\documentclass[aps,prc,superscriptaddress,showpacs,floatfix,reprint]{revtex4-1}
\usepackage[utf8]{inputenc}
\usepackage{amsmath}
\usepackage{color}
\usepackage[frak=esstix]{mathalpha}
\usepackage{makeidx}
\usepackage[english]{babel}
\usepackage{fontenc}
\usepackage{graphicx}
\usepackage{lineno}
\usepackage{enumerate}
\usepackage{booktabs}

\usepackage{amsmath,amssymb}
\usepackage{graphicx}
\usepackage{booktabs}
\usepackage{array}
\usepackage{xcolor}
\usepackage{hyperref}
\usepackage{siunitx}
\usepackage{multirow}
\usepackage{microtype}
\usepackage{float}

\hypersetup{
  colorlinks=true,
  linkcolor=blue,
  citecolor=blue,
  urlcolor=blue
}

\newcommand{\sqrts}{\sqrt{s}}

\newcommand{\pt}{p_{\mathrm{T}}}
\newcommand{\pthat}{\hat{p}_{\mathrm{T}}}
\newcommand{\dphi}{\Delta\phi}
\newcommand{\absdphi}{|\Delta\phi|}
\newcommand{\mjj}{M_{jj}}

\newcommand{\aj}{A_J}
\newcommand{\qhat}{\hat{q}}
\newcommand{\qhatnought}{\hat{q}_0}
\newcommand{\antikt}{anti-$k_T$}

\newcommand{\D}{{\rm d}}

\newcommand{\placeholderfigure}[2]{%
  \begin{center}
  \fbox{\begin{minipage}{0.82\linewidth}
  \centering
  \vspace{0.5em}
  \textbf{Placeholder figure}\par
  #1\par
  \vspace{0.5em}
  \nolinkurl{#2}\par
  \vspace{0.5em}
  \end{minipage}}
  \end{center}
}

\newcommand{\maybeincludegraphics}[2][]{%
  \IfFileExists{#2}{\includegraphics[#1]{#2}}{\placeholderfigure{Put final PDF plot here.}{#2}}
}

\begin{document}

\title{Sensitivity of jet and dijet observables to medium effects in pp collisions at \texorpdfstring{$\sqrts=13$~TeV}{sqrt(s) = 13 TeV} using X-SCAPE/JETSCAPE}

\author{Lakshitha Fernando}
\email{lakshithafernando@wayne.edu}
\affiliation{Department of Physics and Astronomy, Wayne State University, Detroit, Michigan 48201, USA}

\author{Claude A. Pruneau}
\email{claude.pruneau@wayne.edu}
\affiliation{Department of Physics and Astronomy, Wayne State University, Detroit, Michigan 48201, USA}

\author{Joern Putschke}
\email{joern.putschke@wayne.edu}
\affiliation{Department of Physics and Astronomy, Wayne State University, Detroit, Michigan 48201, USA}

\author{Abhijit Majumder}
\email{majumder@wayne.edu}
\affiliation{Department of Physics and Astronomy, Wayne State University, Detroit, Michigan 48201, USA}

\author{Victor Gonzalez}
\affiliation{Department of Physics and Astronomy, Wayne State University, Detroit, Michigan 48201, USA}

\author{Oveis Sheibani}
\affiliation{Department of Physics and Astronomy, Wayne State University, Detroit, Michigan 48201, USA}

\author{Akash Raj}
\affiliation{Department of Physics and Astronomy, Wayne State University, Detroit, Michigan 48201, USA}

\author{Dana Abdallah}
\affiliation{Department of Physics and Astronomy, Wayne State University, Detroit, Michigan 48201, USA}

\date{\today}

\begin{abstract}
Establishing jet quenching in small collision systems, such as high-multiplicity proton--proton collisions at the LHC, is challenging because the expected modifications are small and can be obscured by event-selection effects, multiparton interactions, and multijet production. We examine the response of several jet and dijet observables to medium-induced changes in proton--proton collisions at $\sqrts=13$~TeV using the JETSCAPE/X-SCAPE framework. The study has two parts. First, a simplified Bjorken-expanding brick medium is compared with a vacuum reference for several brick lengths and strengths of transport coefficients. Second, separately generated realistic vacuum and medium samples with fluctuating initial conditions and event-by-event hydrodynamic evolution are compared after selecting the highest-activity 20\% of events. Full \antikt{} jets are reconstructed with $R=0.2$ and $R=0.4$. The observable set comprises dijet invariant mass, leading- and subleading-jet transverse momentum, folded azimuthal separation, dijet asymmetry, Soft-Drop groomed jet radius, and groomed mass. In the brick calculation, the dijet-mass spectrum shows the clearest ordered response: departures from the vacuum reference generally increase with brick length and strength of transport coefficient and are more visible for the narrower jet radius, while the jet-$p_{\rm T}$ and dijet-asymmetry changes are smaller and no clear azimuthal broadening is resolved. In the selected realistic events, the ungroomed medium-to-vacuum ratios remain close to unity overall but contain localized kinematic differences, together with a modest increase of the mean dijet asymmetry. The groomed-radius and groomed-mass ratios likewise show localized differences but do not exhibit a persistent angular narrowing or mass redistribution. These findings quantify the sensitivity of the studied observables to medium-induced jet modifications within the present pp calculations.

\end{abstract}

\maketitle

\section{Introduction}
\label{sec:introduction}

High-multiplicity proton--proton collisions at the LHC exhibit long-range multiparticle azimuthal correlations~\cite{ALICE:2019MultiparticleFlow,ALICE:2021LongRangeJetScale}. A pronounced increase in strange and multi-strange hadron production with charged-particle multiplicity is also measured~\cite{ALICE:2017Strangeness}. These observations have made proton--proton (pp) collisions an important testing ground to seek evidence for collective fluid-like behavior in very small systems. However, their physical origin is yet to be settled. Hydrodynamic expansion can account for part of the measured behavior, but initial-state correlations, color reconnection, interacting strings, and other non-hydrodynamic mechanisms may also contribute~\cite{Noronha:2024SmallSystemsReview}. This motivates the question of whether final-state interactions in high-activity pp events are strong enough to modify energetic parton showers and provide a signature of QGP-like matter.

Jets provide a complementary probe because they originate from large-momentum-transfer partonic scatterings that occur early in the collision~\cite{Majumder:2010qh}. If a short-lived colored medium is formed, the resulting parton showers may interact with it through medium-induced radiation and scattering. Such interactions can in principle affect jet spectra, dijet-mass spectra, dijet momentum balance, and the angular relation between the two jets. In pp collisions, the available medium size and path length are expected to be small, so the resulting changes are expected to be correspondingly small. Quantifying this response, or constraining it, is therefore important for assessing whether these observables have sufficient sensitivity to QGP-like final-state effects in small systems~\cite{Krizek:2025SmallSystemJetQuenching}.

Jet grooming provides a way to reduce sensitivity to soft, wide-angle radiation and to isolate the hard substructure of a jet. In particular, the Soft Drop procedure removes soft branches and identifies a hard splitting characterized by its momentum sharing $z_g$ and angular separation $R_g$~\cite{Larkoski:2014SoftDrop}. Measurements of these groomed observables can then aid in investigating possible modifications of jet substructure due to interactions with the medium. In heavy-ion collisions, ALICE observed a narrowing of the scaled groomed jet radius $\theta_{\rm g}$ in Pb--Pb relative to pp, while finding no significant modification of the groomed momentum-sharing distribution~\cite{ALICE:2022GroomedRadius}. ATLAS subsequently found a strong dependence of jet suppression on the groomed splitting angle, with the widest measured jets substantially more suppressed than the narrowest in central Pb--Pb collisions~\cite{ATLAS:2023SubstructureSuppression}. Groomed-mass measurements also feature sensitivity to the presence of a medium~\cite{Majumder:2014gda}. CMS reported a hint of enhanced probability at large groomed mass in the most central Pb--Pb collisions for part of the measured phase space~\cite{CMS:2018GroomedMass}, whereas a recent ALICE measurement found the groomed jet-mass distribution to be narrower in Pb--Pb than in pp~\cite{ALICE:2025GroomedMass}. These measurements motivate similar investigations in smaller systems, i.e., investigations of whether analogous substructure observables show a resolvable response in a modeled high-activity pp medium.

Recent measurements also show why this search requires care. ALICE observed broader recoil-jet acoplanarity in high-multiplicity pp collisions at $\sqrts=13$ TeV than in minimum-bias events. The same trend appeared in PYTHIA calculations without jet quenching and was traced to a selection bias toward events containing several final-state jets; multiparton interactions were also identified as the dominant background contribution in that analysis~\cite{ALICE:2024HMppAcoplanarity}. Additional partonic activity and multijet final states can therefore alter the azimuthal relation of the selected jets even without jet--medium interactions. Measurements in p--Pb collisions have likewise yielded constraints without a clear quenching signal. ATLAS found jet--hadron correlations consistent with unity within a few percent across the studied centrality range~\cite{ATLAS:2023pPbConstraints}. CMS measured high- to low-multiplicity $x_J$ ratios compatible with simulations that do not include a QGP and placed a 90\% confidence-level upper limit of 1.26\% on the fractional energy loss of the subleading jet~\cite{CMS:2025pPbXJ}. In contrast, a recent measurement in oxygen--oxygen collisions reported suppressed charged-particle production in a light-ion system~\cite{CMS:2025OOSuppression}, further motivating quantitative studies of jet modification in small systems.

JETSCAPE provides a modular framework in which the hard process, shower evolution, medium background, and hadronization can be controlled independently~\cite{Putschke:2020Framework}. It has been applied to both bulk and hard-probe observables in pp and heavy-ion collisions~\cite{JETSCAPE:2024nkj}, including Bayesian constraints on the transport properties of QCD matter~\cite{JETSCAPE:2021MultiSystem} and on the jet transport coefficient $\qhat$~\cite{JETSCAPE:2021Qhat,JETSCAPE:2025QhatBayesian}, as well as multistage descriptions of inclusive jet and hadron suppression~\cite{JETSCAPE:2023InclusiveSuppression}. MATTER describes the high-virtuality shower through virtuality-ordered evolution~\cite{Majumder:2013Matter,Cao:2020Matter}; the PP19 calculation benchmarked the JETSCAPE vacuum setup against pp jet measurements~\cite{JETSCAPE:2020ppResults}, and X-SCAPE extends the framework to correlated soft and hard production in pp and p--A collisions~\cite{Majumder:2023XSCAPEProc,Soudi:2024XSCAPE}.

The ALICE Collaboration reported a measurement of the multiplicity dependence of charged-particle jet production and jet fragmentation in pp collisions at $\sqrts=13$~TeV~\cite{ALICE:2022JetMultiplicity,ALICE:2024IntraJet}; the study was motivated, in part,  by the prospect of extending such measurements to dijet and groomed-jet observables. We use two complementary calculations for pp collisions at $\sqrts=13$~TeV. The first compares a vacuum baseline with a Bjorken-expanding brick in which the configured brick length and transport coefficient are varied. The second compares realistic vacuum and medium samples with fluctuating initial conditions and event-by-event hydrodynamic evolution after a high-activity selection based on charged-particle production from the soft and hard sectors. We study the dijet invariant mass $\mjj$~\cite{JETSCAPE:2020ppResults}, leading- and subleading-jet transverse-momentum spectra, folded azimuthal separation $\absdphi$, and dijet asymmetry $\aj$ for $R=0.2$ and $R=0.4$; the realistic comparison additionally includes the scaled Soft-Drop groomed radius $\theta_{\rm g}=R_{\rm g}/R$~\cite{Larkoski:2014SoftDrop} and groomed mass $m_{\rm g}$~\cite{JETSCAPE:2023hqn,Tachibana:2025rcx}. This common analysis is used to establish the observable response in the controlled calculation and to test whether a corresponding pattern is resolved in the event-by-event small-system calculation.

Section~\ref{sec:framework-setup} describes the two event-generation setups. The jet definition, selections, and observables are given in Sec.~\ref{sec:observables}. Section~\ref{sec:vacbrick-results} presents the controlled vacuum--brick comparison, and Sec.~\ref{sec:realistic-results} gives the realistic pp vacuum--medium results. The main conclusions are summarized in Sec.~\ref{sec:conclusion}, and detailed configuration and statistical information is collected in Appendix~\ref{app:configuration-statistics}.

\section{Framework and event generation}
\label{sec:framework-setup}

The events were produced with the modular JETSCAPE/X-SCAPE framework~\cite{Putschke:2020Framework,Soudi:2024XSCAPE}. Two matched comparisons were performed: a controlled vacuum--brick benchmark and a realistic event-by-event vacuum--medium calculation. Each vacuum sample serves only as the reference for its corresponding medium setup. Both studies simulate pp collisions at $\sqrts=13$~TeV with hard scatterings in the interval $30<\pthat<3000$~GeV.

\subsection{Controlled vacuum--brick study}
\label{subsec:vacbrick-config}

PYTHIA~8 was used to generate the hard scattering~\cite{Sjostrand:2015Pythia8,Bierlich:2022Pythia83}, while MATTER generated the outgoing final-state shower~\cite{Majumder:2013Matter,Cao:2020Matter}. The vacuum and brick samples share the hard-process, shower, and hadronization settings, which follow the core PP19 prescription of Kumar et al.~\cite{JETSCAPE:2020ppResults}; the complete settings are listed in Table~\ref{tab:controlled-parameters}.

For the vacuum reference, MATTER evolved without coupling to a medium. In the brick calculation, the shower propagated through a transversely uniform background undergoing Bjorken longitudinal expansion. The medium began at proper time $\tau_0=0.2$~fm/$c$ with temperature $T_0=0.35$~GeV and cooled according to
\begin{equation}
 T(\tau)=T_0\left(\frac{\tau_0}{\tau}\right)^{1/3},
 \label{eq:brick-temperature}
\end{equation}
where $\tau$ denotes the  proper time. This temperature profile follows boost-invariant Bjorken expansion~\cite{Bjorken:1983BoostInvariant} and is the behavior implemented by the expanding-brick option in JETSCAPE~\cite{Putschke:2020Framework}. Uniform media of this type are commonly used to isolate the response of shower models to controlled medium scales~\cite{JETSCAPE:2017StaticMedium}.

The jet transport coefficient is physically characterized by the transverse-momentum broadening acquired by an energetic parton per unit propagation length~\cite{JETSCAPE:2021Qhat},
\begin{equation}
 \qhat \equiv \frac{d\langle k_\perp^2\rangle}{d\ell},
 \label{eq:qhat-definition}
\end{equation}
where $k_\perp$ is the transverse momentum transferred to the propagating parton and $\ell$ is distance along its trajectory. The controlled scan uses configured brick lengths $L=0.5$, 1.0, and 2.0~fm and constant MATTER transport coefficients $\qhat_0=0.25$ and 0.50~GeV$^2$/fm. The brick length $L$ is the maximum geometric length available to the shower; the realized interaction distance can be shorter because the medium begins at $\tau_0$, cools during the evolution, and ceases to affect the shower below the medium cutoff temperature. For context, earlier JETSCAPE static-medium studies used brick lengths of 4 and 8~fm~\cite{JETSCAPE:2017StaticMedium}, while phenomenological analyses of central heavy-ion collisions extracted local transport coefficients of approximately $\qhat=1.2\pm0.3$~GeV$^2$/fm for a 10~GeV quark at $\tau_0=0.6$~fm/$c$ in $\sqrt{s_{NN}}=200$~GeV Au--Au collisions and $\qhat=1.9\pm0.7$~GeV$^2$/fm in $\sqrt{s_{NN}}=2.76$~TeV Pb--Pb collisions~\cite{JET:2014Qhat}.

Medium recoil and its associated backreaction are included in the brick calculation and absent from the vacuum reference; their treatment in the jet reconstruction is described in Sec.~\ref{sec:observables}. The vacuum sample and every reported brick configuration contain $10^7$ generated events.

\subsection{Realistic vacuum--medium study}
\label{subsec:realistic-workflow}

The realistic calculation is based on the X-SCAPE soft--hard framework of Soudi et al.~\cite{Soudi:2024XSCAPE}. PYTHIA generates the hard scattering with its own initial-state shower disabled, and the initial-state radiation is generated by iMATTER through backward virtuality-ordered evolution of the incoming partons. A three-dimensional Monte Carlo Glauber model supplies the fluctuating initial condition from the energy remaining after the hard partons are removed, MUSIC evolves the soft medium, MATTER generates the final-state shower, and iSS samples hadrons from the freeze-out hypersurface. The hard-sector partons are hadronized with the X-SCAPE color-connected string hadronization, in which the final-state partons are color-connected to the initial-state partons and the Glauber beam remnants. Complete numerical settings are listed in Tables~\ref{tab:realistic-parameters-a} and~\ref{tab:realistic-parameters-b}.

The hard-process configuration is common to the realistic vacuum and medium samples. PYTHIA phase-space biasing was used to improve the population of the steeply falling high-$\pthat$ region~\cite{Sjostrand:2015Pythia8,Bierlich:2022Pythia83}. Every histogram entry, as well as the event-activity ranking described in Sec.~\ref{sec:observables}, carries the corresponding compensating PYTHIA event weight, so that weighted distributions reproduce the unbiased hard-process cross section. The parent samples are therefore hard-process selected rather than minimum-bias pp samples. Each realistic sample contains $10^6$ generated hard/hydrodynamic events; generating the two realistic samples required approximately six weeks on the Wayne State Grid, which illustrates the computational cost of the event-by-event soft--hard calculation.

The soft sector was initialized with the three-dimensional Glauber model and evolved with MUSIC from $\tau_0=0.6$~fm/$c$~\cite{Soudi:2024XSCAPE,Shen:2020Geometry3D,Shen:2018DynamicalInitialState,Shen:2022Longitudinal,Schenke:2010MUSIC}. The vacuum and medium samples use the same initial-state prescription, equation of state, viscous parameters, freeze-out criterion, and iSS settings. At freeze-out, iSS performs Cooper--Frye sampling of the hydrodynamic hypersurface~\cite{Cooper:1974FreezeOut,Shen:2016iEBEVISHNU}, with resonance decays enabled.

The distinction between the two realistic samples lies in the treatment of the hard shower. In the vacuum control, the fluctuating initial state, hydrodynamic evolution, freeze-out surface, and soft hadrons were generated, but the hydrodynamic fields were not supplied to MATTER. In the medium sample, the MUSIC history was passed to MATTER, allowing the shower to interact with fluid cells above $T_c=0.16$~GeV. The remaining shower and hadronization settings were matched between the two samples.

\section{Jet reconstruction and observables}
\label{sec:observables}

Charged and neutral final-state hadrons are clustered into full jets with the \antikt{} algorithm implemented in FastJet~\cite{Cacciari:2008AntiKt,Cacciari:2012FastJet}. In the controlled brick samples, recoil hadrons are retained as jet constituents, while status $-1$ backreaction hadrons enter the clustering with negative four-momentum, so that they are subtracted in the $E$-scheme recombination; reconstructed clusters with net negative energy are discarded before the jet selection. This is a clustering-level implementation of the recoil--hole subtraction used in JETSCAPE calculations~\cite{JETSCAPE:2023InclusiveSuppression}. In the realistic samples, iSS-sampled soft hadrons enter the event-activity estimator but are excluded before jet clustering, implementing the perfect soft-background subtraction assumed in the X-SCAPE jet analysis~\cite{Soudi:2024XSCAPE}. Jet constituents are required to satisfy $\pt\geq0.15$~GeV and $|\eta|<3.0$. Jets are reconstructed with $R=0.2$ and $R=0.4$, and accepted jets satisfy $\pt^{\rm jet}\geq10$~GeV and
\begin{equation}
 |\eta_{\rm jet}|<3.0-R,
 \label{eq:jet-acceptance}
\end{equation}
so that the complete jet cone remains inside the constituent acceptance. The two jet radii are used to examine how much of the redistributed shower energy is recovered by a wider cone.

For the realistic samples, event activity is defined from charged-particle production in the soft and hard sectors. The iSS configuration produces $N_{\rm iSS}=1000$ repeated Cooper--Frye samples of each hydrodynamic freeze-out hypersurface. In the event-activity estimator used in this work, the accumulated charged status-11 yield is divided by $N_{\rm iSS}$ to obtain the mean soft charged multiplicity per hydrodynamic event before adding the charged hard-sector yield,
\begin{equation}
 N_{\rm ch}^{\rm EA}=\frac{N_{\rm ch}^{\rm soft}}{N_{\rm iSS}}+N_{\rm ch}^{\rm hard},
 \qquad N_{\rm iSS}=1000.
 \label{eq:event-activity-estimator}
\end{equation}
Such repeated particlization sampling of a fixed hypersurface is commonly used to obtain multiple hadronic realizations from the same hydrodynamic event~\cite{Shen:2016iEBEVISHNU,McNelis:2021iS3D}; the combined soft--hard construction follows the event structure of X-SCAPE~\cite{Soudi:2024XSCAPE}. The activity count uses all charged hadrons in the event record without an additional $\eta$ or $\pt$ requirement. The highest-activity 20\% of the weighted parent distribution is retained, hereafter the top-20\% event-activity (EA) class, with the threshold determined separately for the vacuum and medium samples. Because the estimator includes the hard-sector yield, which the medium can modify, the selected event classes need not be identical in the two samples. This defines a model-level event-activity class and is not intended to reproduce a specific detector multiplicity estimator.

An event entered the dijet analysis when it contained at least two accepted jets. The two jets with the largest transverse momenta were labeled leading and subleading; additional accepted jets were not vetoed. For the full-range dijet selection, the leading jet was restricted to $10\leq\pt^{\rm lead}\leq500$~GeV. The dijet-selected leading- and subleading-jet spectra are reported as the differential cross sections $\D\sigma/\D\pt^{\rm lead}$ and $\D\sigma/\D\pt^{\rm sublead}$. These distributions are sensitive to migration between transverse-momentum bins caused by shower--medium interactions.

The azimuthal separation of the leading and subleading jets was folded into $0\leq\absdphi\leq\pi$ according to
\begin{equation}
 \absdphi=\arccos\!\left[\cos(\phi_1-\phi_2)\right],
 \label{eq:dphi-folded}
\end{equation}
where $\phi_1$ and $\phi_2$ are the azimuthal angles of the leading and subleading jets, respectively. The full $\absdphi$ distribution was filled without an additional back-to-back requirement. A medium effect could alter the height or width of the peak near $\pi$, although acoplanarity in pp collisions can also be modified by multijet production and event selection~\cite{ALICE:2024HMppAcoplanarity}.

The dijet asymmetry is defined by
\begin{equation}
 \aj=\frac{\pt^{\rm lead}-\pt^{\rm sublead}}
 {\pt^{\rm lead}+\pt^{\rm sublead}}.
 \label{eq:aj-definition}
\end{equation}
A balanced dijet has $\aj\simeq0$. The $\aj$ distributions are evaluated only for pairs that satisfy
\begin{equation}
 \pi-\absdphi<30^{\circ}.
 \label{eq:back-to-back}
\end{equation}

The dijet invariant mass was obtained from the reconstructed jet four-momenta,
\begin{equation}
 \mjj^2=(p_1+p_2)^2,
 \label{eq:mjj-fourvector}
\end{equation}
where $p_i^{\mu}=(E_i,\mathbf{p}_i)$ is the four-momentum of jet $i$. Expressed in terms of the jet masses ($m_i$), transverse momenta ($p_{\mathrm T,i}$), rapidities ($y_i$), and azimuthal angles ($\phi_i$), this becomes
\begin{widetext}
\begin{equation}
 \begin{aligned}
 \mjj^2={}&m_1^2+m_2^2+2\left[m_{\mathrm T,1}m_{\mathrm T,2}\cosh(y_1-y_2)
 -p_{\mathrm T,1}p_{\mathrm T,2}\cos(\phi_1-\phi_2)\right],
 \end{aligned}
 \label{eq:mjj-expanded}
\end{equation}
\end{widetext}
with
\begin{equation}
 m_{\mathrm T,i}=\sqrt{m_i^2+p_{\mathrm T,i}^2}.
 \label{eq:jet-transverse-mass}
\end{equation}
When the reconstructed jet masses are small compared with their transverse momenta, Eq.~\eqref{eq:mjj-expanded} reduces to
\begin{equation}
 \mjj^2\simeq2p_{\mathrm T,1}p_{\mathrm T,2}
 \left[\cosh(\Delta y)-\cos(\dphi)\right],
 \label{eq:mjj-massless}
\end{equation}
where $\Delta y=y_1-y_2$ and $\dphi=\phi_1-\phi_2$. The $\mjj$ distribution uses the same 30$^{\circ}$ back-to-back requirement and is displayed from 60 to 3000~GeV with logarithmic binning.

The internal structure of inclusive jets in the realistic vacuum and medium samples is studied with the Soft-Drop grooming procedure~\cite{Larkoski:2014SoftDrop}. After the parent jet is reconstructed with the \antikt{} algorithm, its constituents are reclustered with the Cambridge--Aachen algorithm so that the declustering sequence follows the angular structure of the shower. At each splitting into branches 1 and 2, we define
\begin{equation}
 z=\frac{\min\!\left(p_{\mathrm T,1},p_{\mathrm T,2}\right)}
 {p_{\mathrm T,1}+p_{\mathrm T,2}},
 \qquad
 \theta=\frac{\Delta R_{12}}{R},
 \label{eq:softdrop-z-theta}
\end{equation}
where
\begin{equation}
 \Delta R_{12}=\sqrt{(\Delta y_{12})^2+(\Delta\phi_{12})^2},
 \label{eq:softdrop-dr}
\end{equation}
where $p_{\mathrm T,i}$ is the transverse momentum of branch $i$, $\Delta y_{12}$ and $\Delta\phi_{12}$ are the rapidity and azimuthal separations of the branches, and $R$ is the parent-jet resolution parameter. The first declustering that satisfies
\begin{equation}
 z>z_{\rm cut}\,\theta^{\beta}
 \label{eq:softdrop-condition}
\end{equation}
defines the tagged hard splitting. The results shown in this paper use
\begin{equation}
 z_{\rm cut}=0.1,\qquad \beta=0,
 \label{eq:softdrop-settings}
\end{equation}
a commonly used Soft-Drop configuration in pp and heavy-ion jet-substructure studies~\cite{Larkoski:2014SoftDrop,CMS:2018GroomedMass}.

Two groomed observables are reported. The groomed radius is the angular separation of the tagged branches,
\begin{equation}
 R_{\rm g}=\Delta R_{12},
 \qquad
 \theta_{\rm g}=\frac{R_{\rm g}}{R},
 \label{eq:theta-g-definition}
\end{equation}
so that $\theta_{\rm g}$ measures the opening angle of the hard splitting relative to the parent-jet radius. The groomed jet mass is obtained from the four-momentum $p_{\rm g}^{\mu}$ of the Soft-Drop groomed jet,
\begin{equation}
 m_{\rm g}^{2}=p_{\rm g}^{\mu}p_{{\rm g},\mu}.
 \label{eq:mg-definition}
\end{equation}
Only jets containing a splitting that passes Eq.~\eqref{eq:softdrop-condition} contribute to the groomed distributions. The displayed results use inclusive parent jets with $60<\pt^{\rm jet}<80$~GeV and $R=0.2$.

The groomed spectra are normalized to the weighted inclusive parent-jet yield in the same jet-radius and parent-$\pt$ interval. For $X\in\{\theta_{\rm g},m_{\rm g}\}$ and histogram bin $k$,
\begin{equation}
 \left.\frac{1}{N_{\rm jet,inc}}\frac{dN_{\rm gr}}{dX}\right|_k
 =\frac{1}{\Delta X_k}
 \frac{\displaystyle\sum_{j\in k,\,{\rm tagged}}\omega_{e(j)}}
 {\displaystyle\sum_{j\in {\rm inclusive}}\omega_{e(j)}},
 \label{eq:groomed-normalization}
\end{equation}
where $\omega_{e(j)}$ is the compensating PYTHIA weight of the event containing jet $j$. Because the distributions are normalized to the inclusive parent-jet yield rather than to the tagged-jet yield, their integral over the complete histogram range equals the weighted Soft-Drop tagging fraction $f_{\rm tagged}$, quoted in the groomed-jet figure legends. The statistical treatment retains the correlation between the tagged numerator and inclusive denominator, as summarized in Appendix~\ref{app:statistical-conventions}.

The ungroomed jet and dijet observables are presented as absolute, cross-section-weighted distributions. For a generated sample with cross section $\sigma_{\rm gen}$ and per-event sampling weights $\omega_i$, the physical weight assigned to event $i$ is
\begin{equation}
 w_i=\sigma_{\rm gen}\frac{\omega_i}{\sum_{j=1}^{N_{\rm norm}}\omega_j},
 \label{eq:event-weight}
\end{equation}
where the normalization sum runs over the full generated parent sample. The controlled samples use $\omega_i=1$, while the realistic samples use the compensating PYTHIA weights. Event activity determines which realistic events contribute, while the normalization remains that of the full parent sample. For a histogram bin $k$ of width $\Delta X_k$, the differential cross section is
\begin{equation}
 \left.\frac{d\sigma}{dX}\right|_k
 =\frac{1}{\Delta X_k}\sum_{i\in k,\,i\in {\rm selected}}w_i,
 \label{eq:differential-cross-section}
\end{equation}
with the selection condition omitted for the controlled samples. Statistical uncertainties of the absolute spectra include the cross-section-normalization uncertainty; ratio uncertainties are propagated as described in Appendix~\ref{app:statistical-conventions}.

\section{Vacuum and controlled brick results}
\label{sec:vacbrick-results}

\begin{figure*}[!htbp]
  \centering
  \includegraphics[width=0.4900\linewidth]{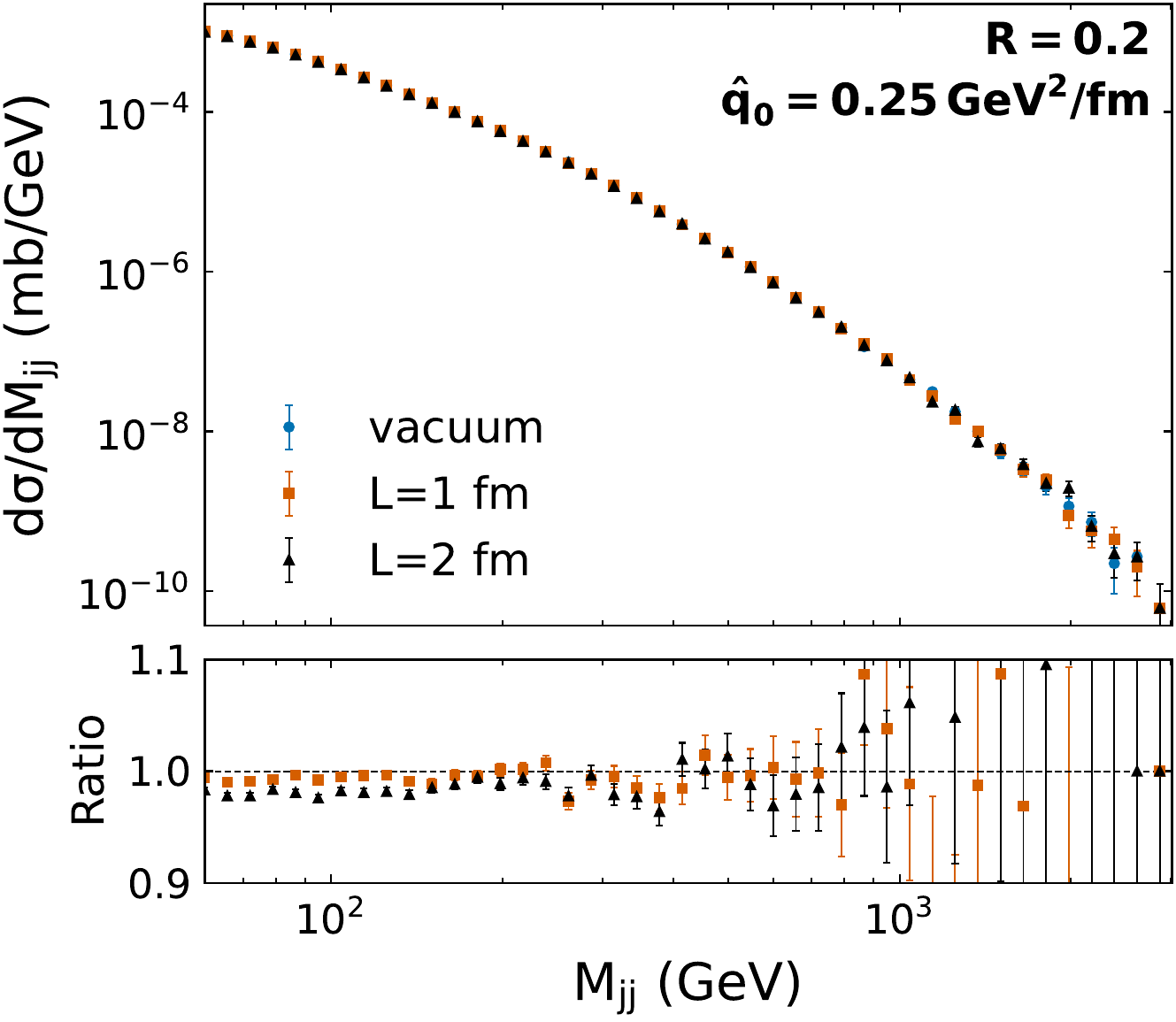}
  \includegraphics[width=0.4900\linewidth]{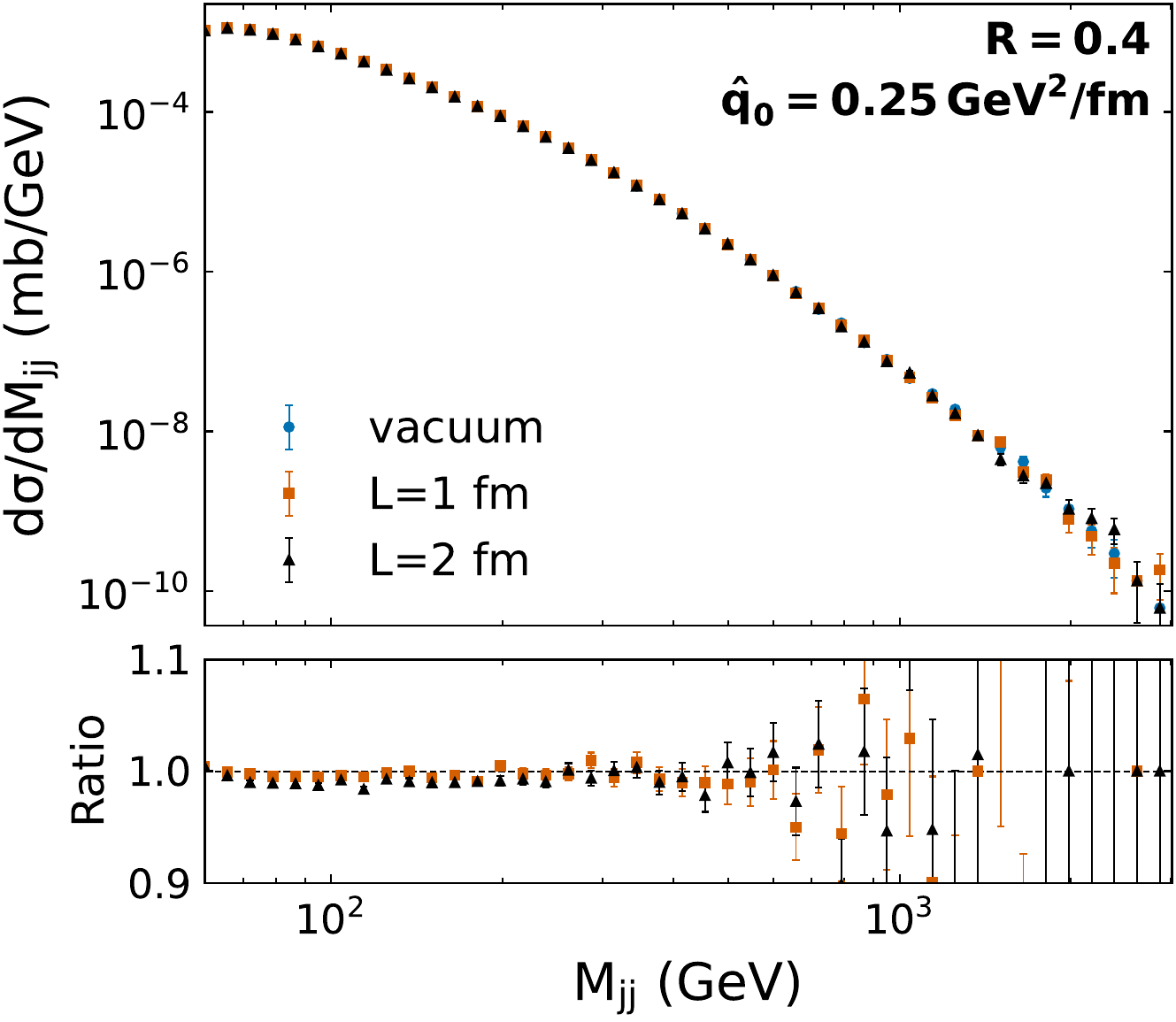}\\[0.5em]
  \includegraphics[width=0.4900\linewidth]{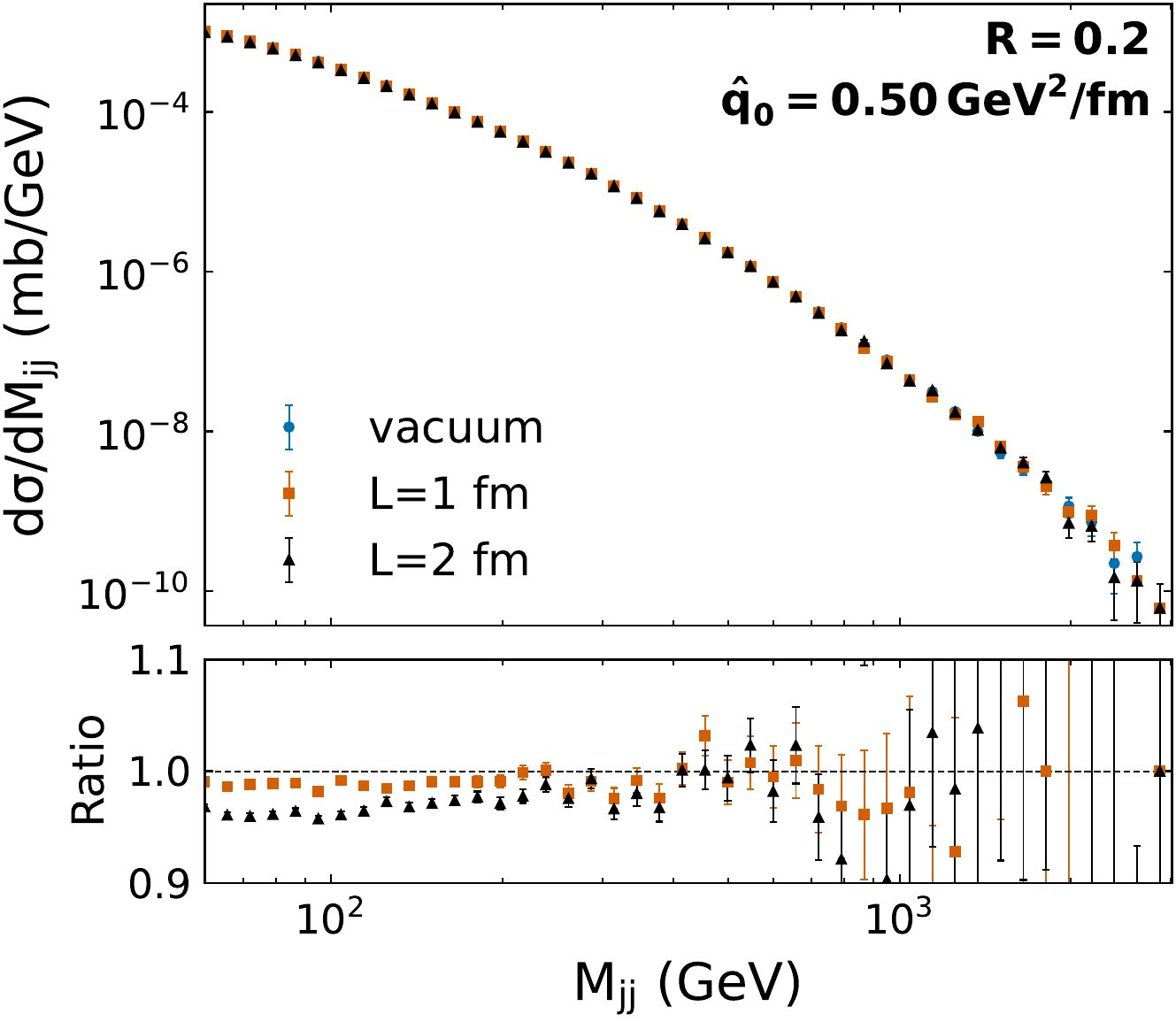}
  \includegraphics[width=0.4900\linewidth]{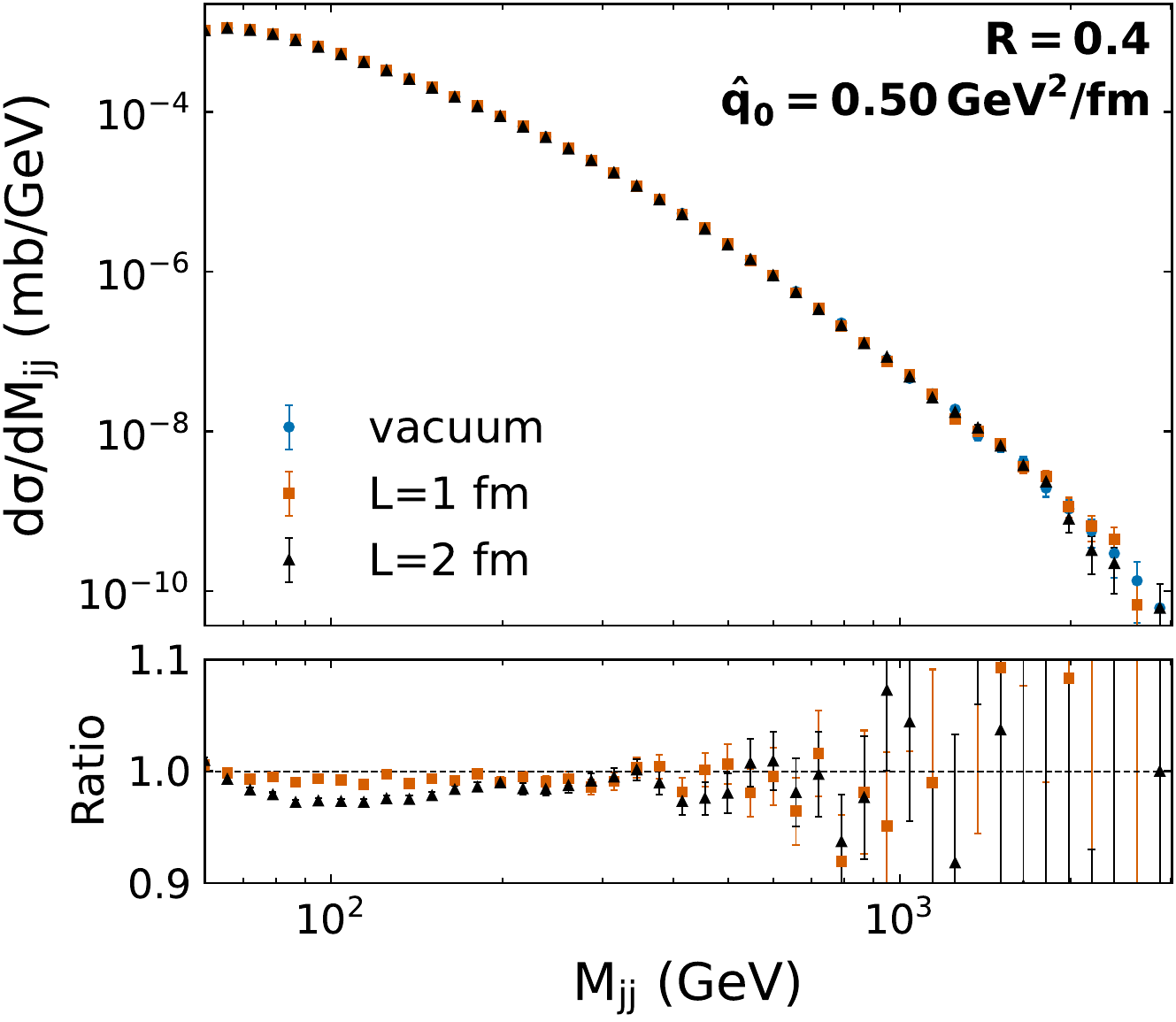}
  \caption{Dijet invariant-mass distributions for the vacuum and brick calculations with $10\leq\pt^{\rm lead}\leq500$ GeV. The upper row shows $\qhatnought=0.25$~GeV$^2$/fm and the lower row shows $\qhatnought=0.50$~GeV$^2$/fm. Jets are reconstructed with $R=0.2$ in the left column and $R=0.4$ in the right column. The displayed brick curves correspond to $L=1.0$ and $2.0$~fm. The lower panels show brick divided by vacuum.}
  \label{fig:vacbrick-mjj}
\end{figure*}

Figure~\ref{fig:vacbrick-mjj} presents the dijet-mass spectra and the corresponding brick-to-vacuum ratios for $10\leq\pt^{\rm lead}\leq500$ GeV. The $L=0.5$~fm results are omitted because they are nearly indistinguishable from the vacuum reference in the displayed range. The mass observable is shown for both transport strengths in order to display the ordering with $\qhatnought$ directly. In the statistically well-populated low- and intermediate-mass region, the brick-to-vacuum ratio is generally below unity and the departure from the vacuum reference becomes more pronounced as the configured brick length increases. Increasing $\qhatnought$ from 0.25 to 0.50~GeV$^2$/fm strengthens this ordering, most clearly for $R=0.2$. The $R=0.4$ ratios remain closer to unity overall, while the highest-mass bins have larger statistical uncertainties and do not support the same level of comparison. This radius dependence is qualitatively consistent with less recovery of redistributed shower momentum in a narrower jet cone, although the dijet mass also depends on the momenta and relative geometry of both jets and the figure alone does not isolate the underlying mechanism.

\begin{figure*}[!htbp]
  \centering
  \includegraphics[width=0.4900\linewidth]{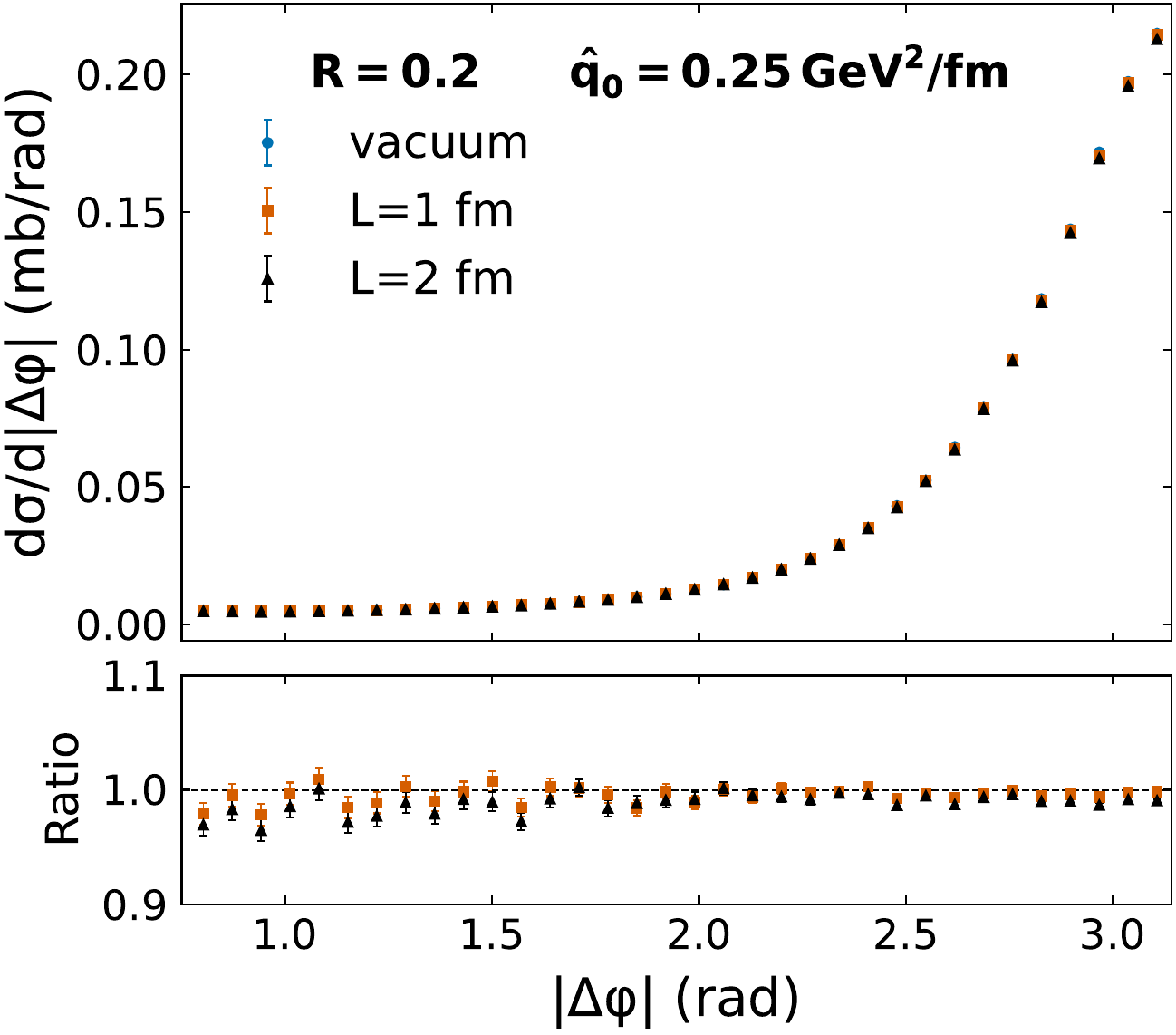}
  \includegraphics[width=0.4900\linewidth]{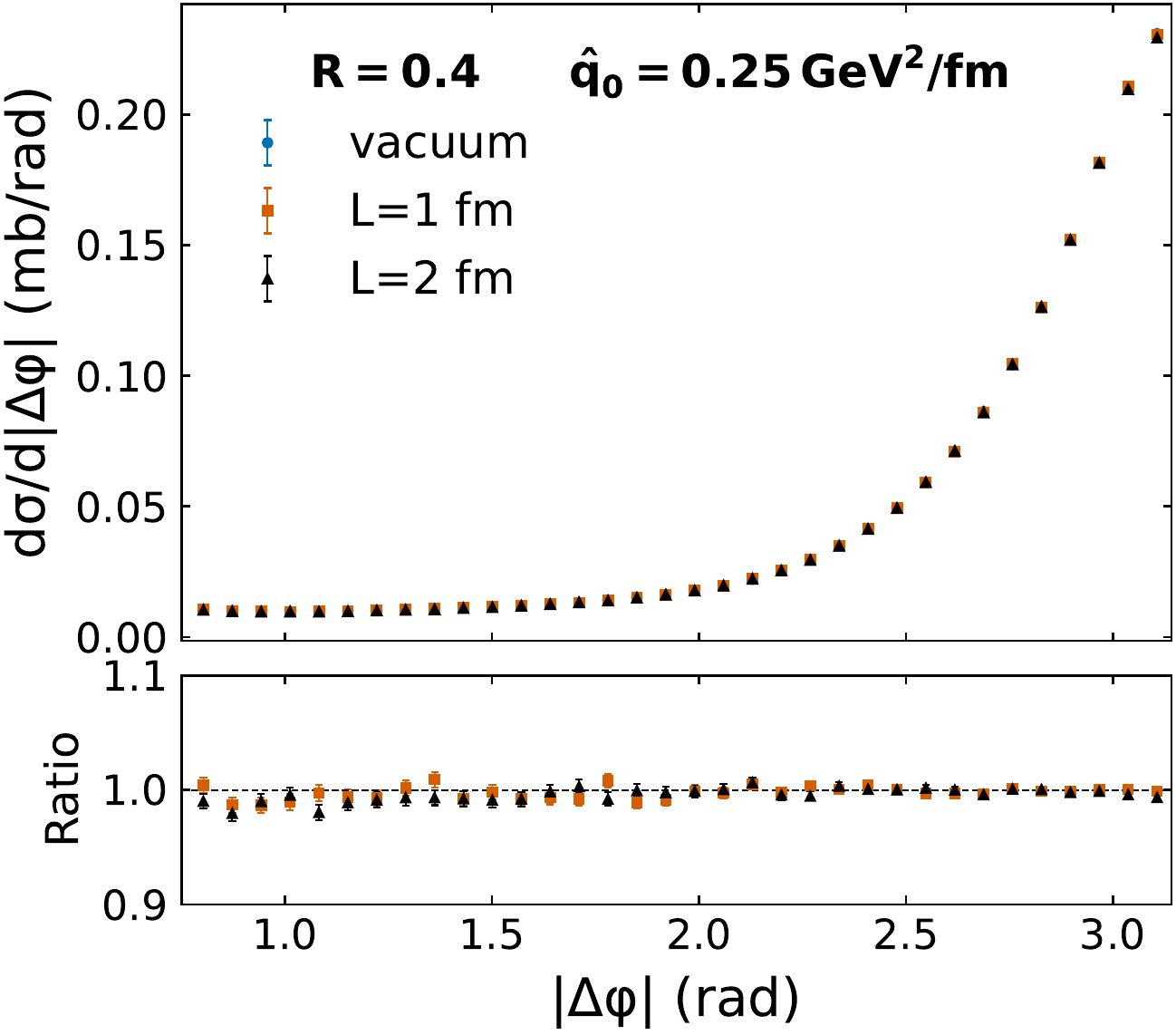}
  \caption{Folded azimuthal-separation distributions for the vacuum and brick samples with $10\leq\pt^{\rm lead}\leq500$ GeV and $\qhatnought=0.25$~GeV$^2$/fm. The left panel shows jets reconstructed with $R=0.2$, and the right panel shows $R=0.4$. Curves and ratio panels as in Fig.~\ref{fig:vacbrick-mjj}.}
  \label{fig:vacbrick-dphi}
\end{figure*}

The folded azimuthal-separation distributions at $\qhatnought=0.25$~GeV$^2$/fm are shown in Fig.~\ref{fig:vacbrick-dphi}. Vacuum and brick curves have nearly the same overall shape for both radii. For $R=0.2$, the $L=2$~fm ratio lies slightly below unity near the back-to-back peak and somewhat further below unity at smaller $\absdphi$, where the absolute yield is much smaller, so the acoplanar region is depleted rather than enhanced. The $R=0.4$ ratios remain closer to unity over the well-populated region. Importantly, the comparison does not show a coherent enhancement away from $\absdphi=\pi$ accompanying a depletion of the back-to-back peak; within the present statistical precision, no clear medium-induced broadening pattern is resolved.

\begin{figure*}[!htbp]
  \centering
  \includegraphics[width=0.4900\linewidth]{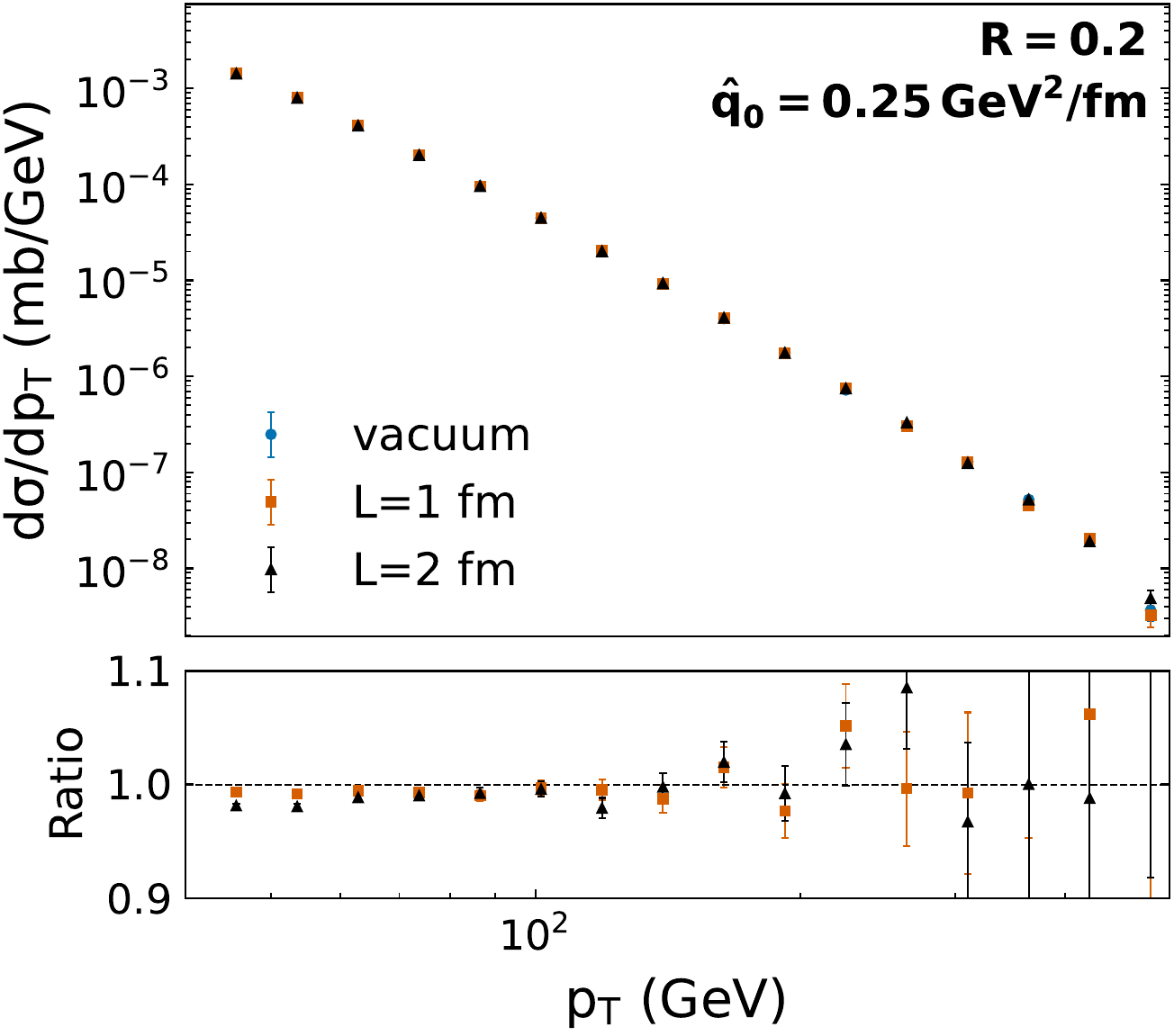}
  \includegraphics[width=0.4900\linewidth]{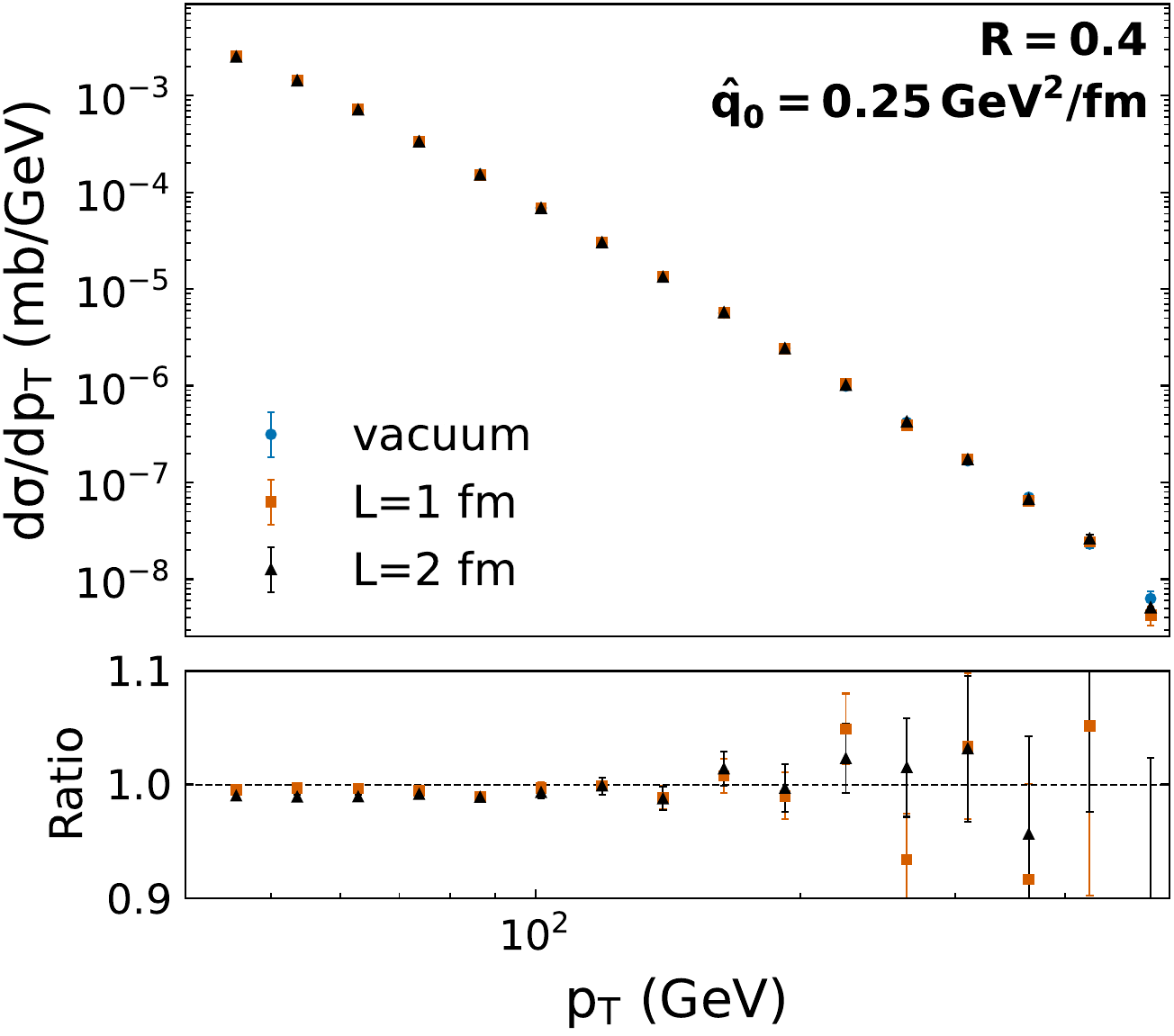}\\[0.5em]
  \includegraphics[width=0.4900\linewidth]{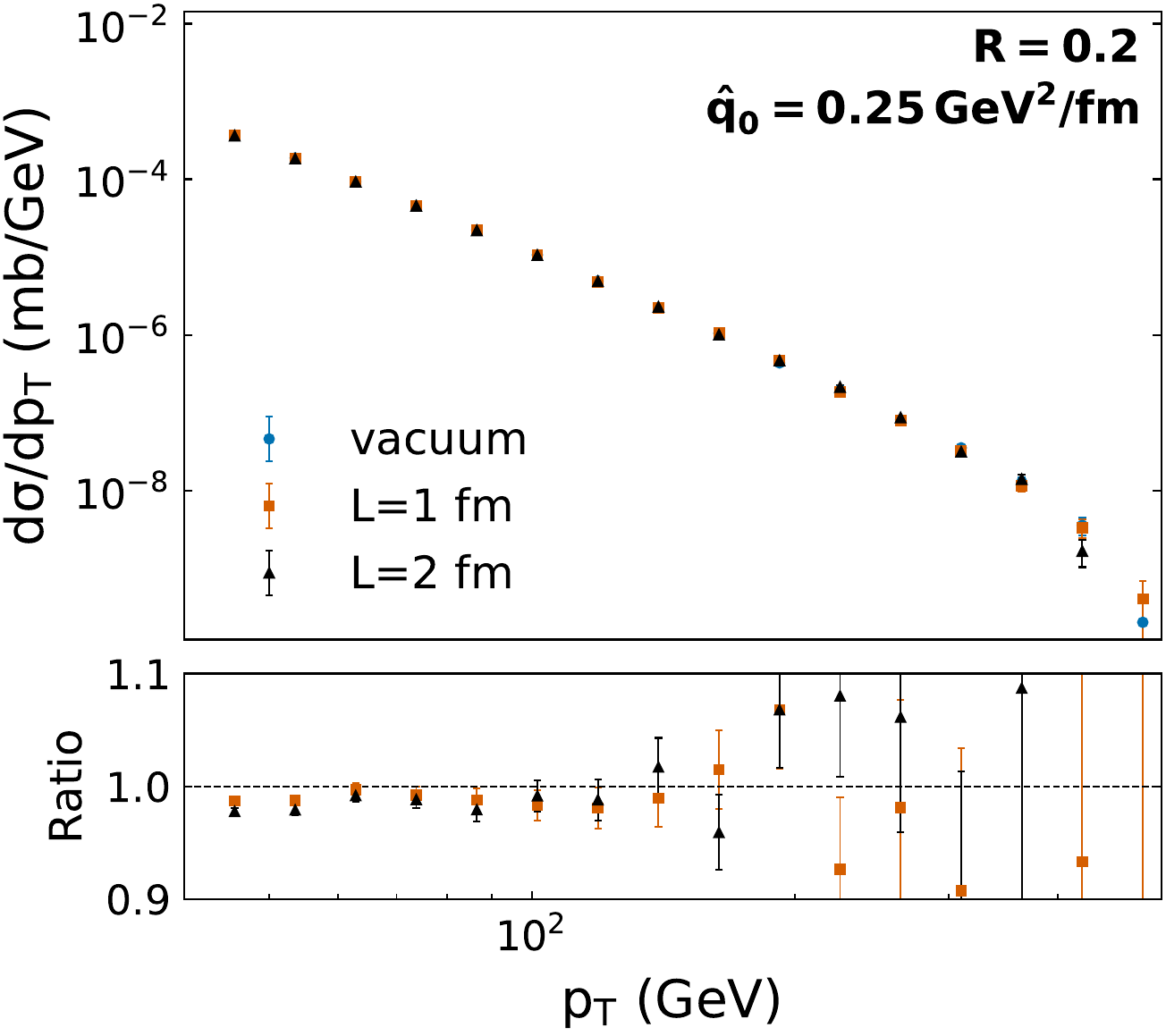}
  \includegraphics[width=0.4900\linewidth]{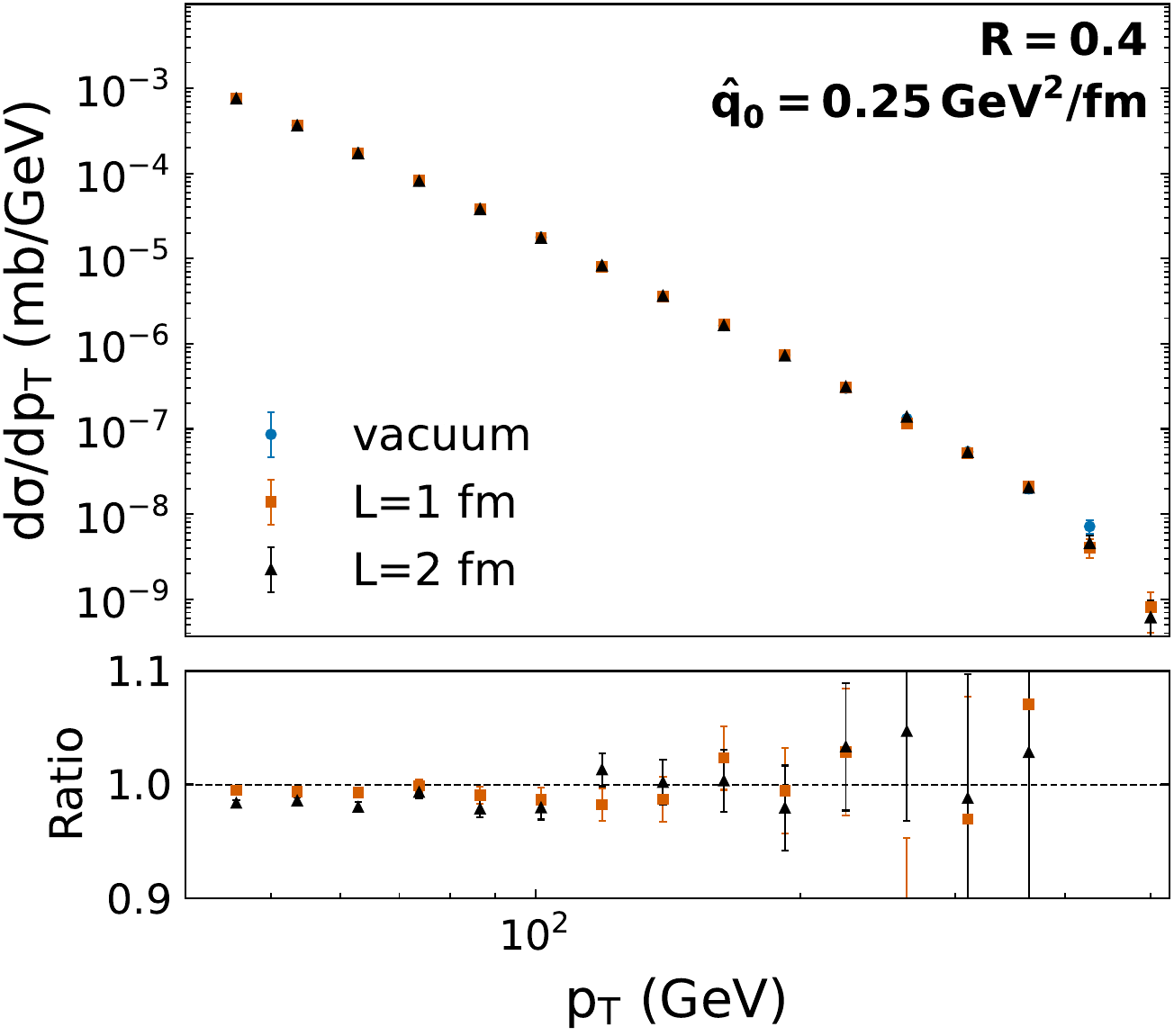}
  \caption{Dijet-selected leading-jet spectra (upper row) and subleading-jet spectra (lower row) for the vacuum and brick samples at $\qhatnought=0.25$~GeV$^2$/fm. Results with $R=0.2$ are shown in the left column and results with $R=0.4$ in the right column. Curves and ratio panels as in Fig.~\ref{fig:vacbrick-mjj}.}
  \label{fig:vacbrick-jetpt}
\end{figure*}

Figure~\ref{fig:vacbrick-jetpt} compares the leading- and subleading-jet spectra at $\qhatnought=0.25$~GeV$^2$/fm. In the best-populated part of the displayed range, the $R=0.2$ spectra for the longer bricks tend to lie below the vacuum reference, while the $R=0.4$ ratios are generally closer to unity. The subleading-jet spectrum exhibits a somewhat stronger downward trend than the leading-jet spectrum for the longest brick. At high $p_{\rm T}$ the statistical uncertainties increase and the ratio ordering becomes less informative. The observed radius dependence is compatible with modest momentum migration outside the smaller jet cone, but the spectra alone do not determine the detailed redistribution mechanism.

\begin{figure*}[!htbp]
  \centering
  \includegraphics[width=0.4851\linewidth]{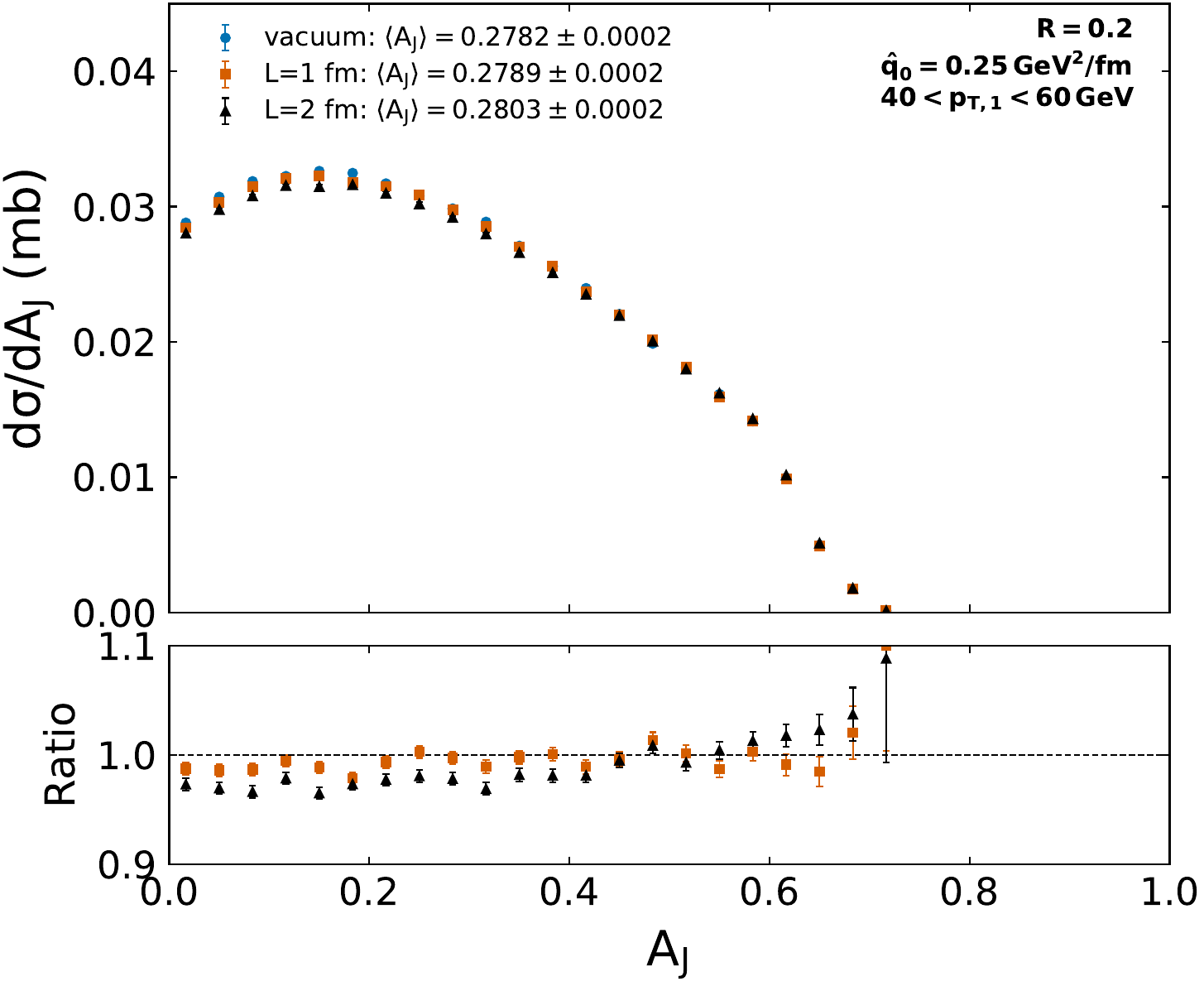}
  \includegraphics[width=0.4949\linewidth]{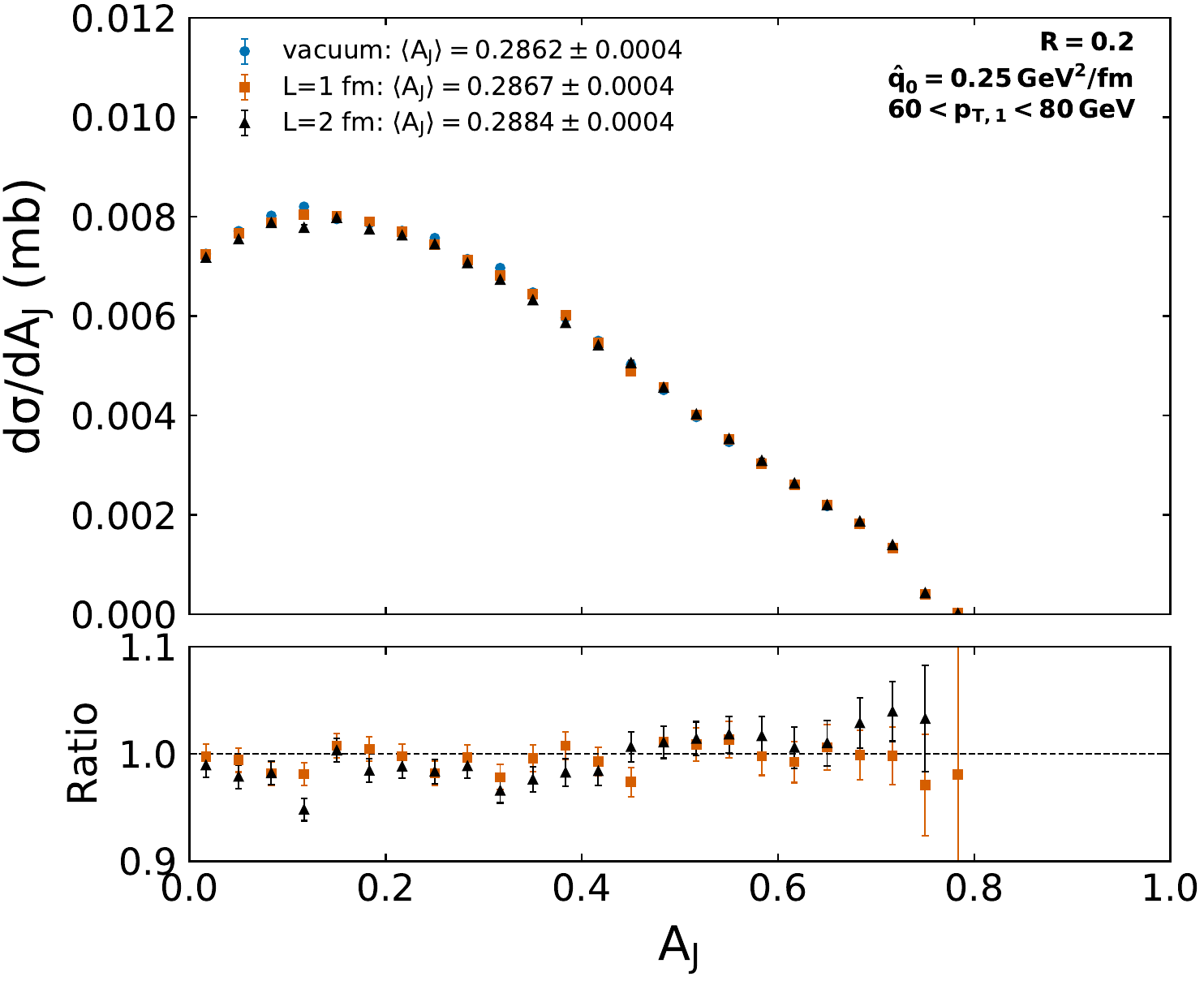}
  \caption{Dijet-asymmetry distributions for the vacuum and brick calculations with $R=0.2$ and $\qhatnought=0.25$~GeV$^2$/fm. The left panel covers $40<\pt^{\rm lead}<60$ GeV and the right panel covers $60<\pt^{\rm lead}<80$ GeV. Curves and ratio panels as in Fig.~\ref{fig:vacbrick-mjj}.}
  \label{fig:vacbrick-aj}
\end{figure*}

The dijet-asymmetry distributions in Fig.~\ref{fig:vacbrick-aj} use $R=0.2$ and $\qhatnought=0.25$~GeV$^2$/fm. For $40<\pt^{\rm lead}<60$ GeV, the mean changes from $0.2782\pm0.0002$ in vacuum to $0.2803\pm0.0002$ for $L=2$ fm, an increase of $0.0021$. For $60<\pt^{\rm lead}<80$ GeV, the mean changes from $0.2862\pm0.0004$ in vacuum to $0.2884\pm0.0004$ for $L=2$ fm, an increase of $0.0022$. In both intervals the displayed $L=1$ and 2 fm distributions overlap strongly with the vacuum reference, so the change in the overall $\aj$ shape remains modest; the $L=2$~fm ratio lies slightly below unity at small $\aj$ and above unity at large $\aj$, consistent with the small increase of the mean.

\section{Realistic pp vacuum--medium comparison}
\label{sec:realistic-results}

\begin{figure*}[!htbp]
  \centering
  \includegraphics[width=0.4900\linewidth]{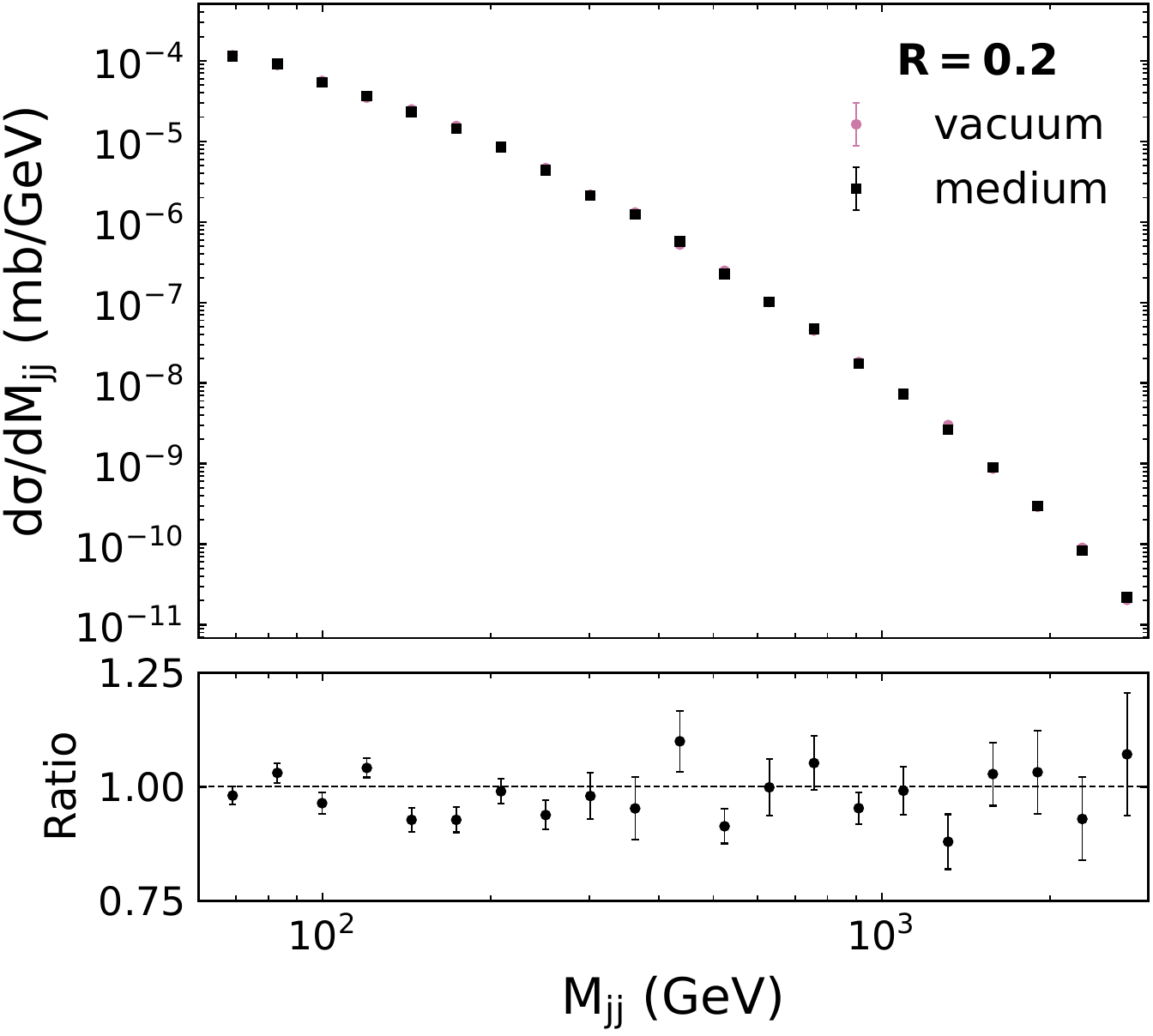}
  \includegraphics[width=0.4900\linewidth]{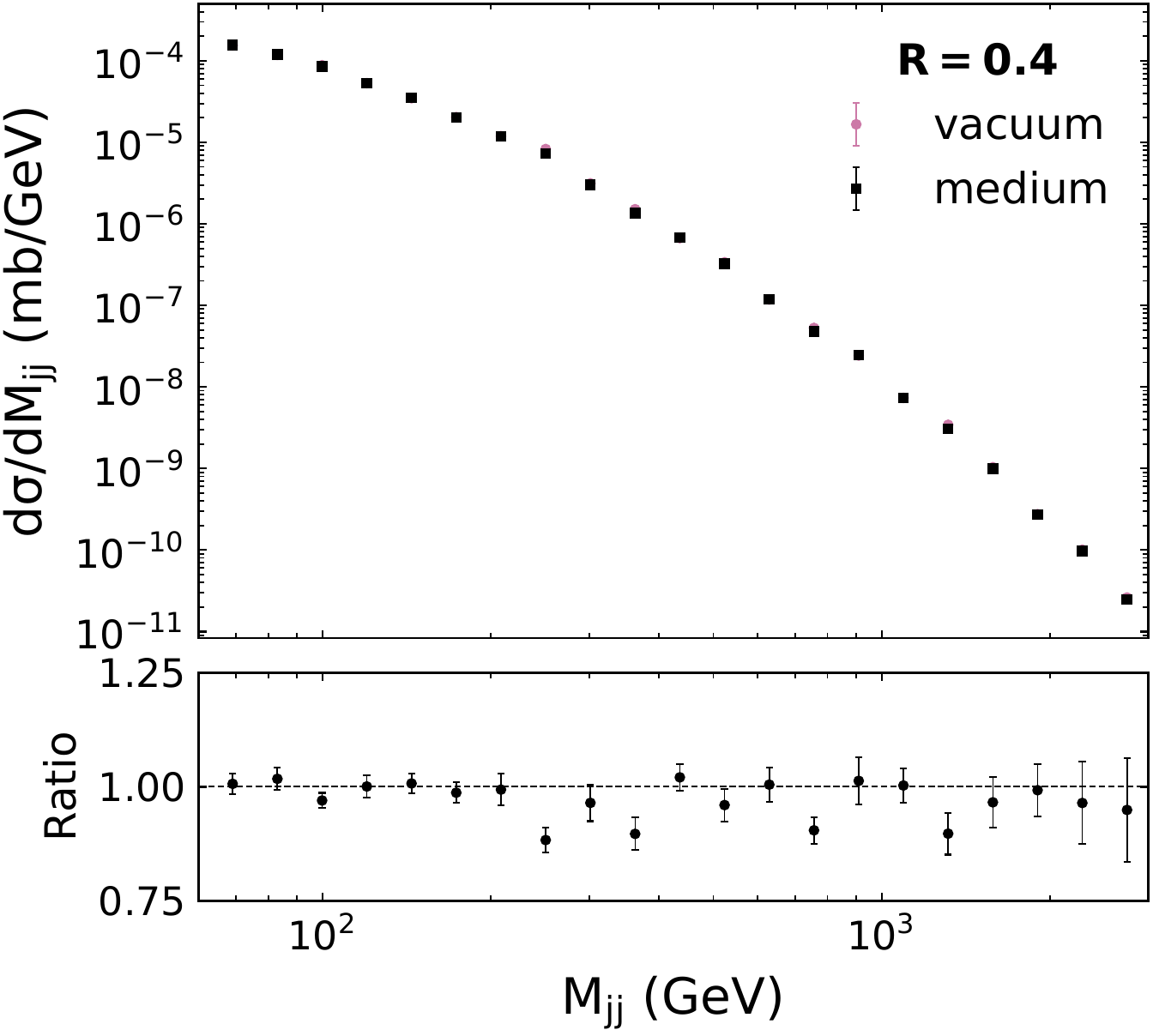}
  \caption{Dijet invariant-mass distributions for the realistic vacuum and medium calculations in the top-20\% EA class, using $R=0.2$ (left) and $R=0.4$ (right). The lower panels show medium divided by vacuum.}
  \label{fig:vacmedium-mjj}
\end{figure*}

Figure~\ref{fig:vacmedium-mjj} shows the dijet invariant-mass spectra for $R=0.2$ and $R=0.4$. The vacuum and medium spectra remain close over many orders of magnitude. The medium-to-vacuum ratios lie mostly slightly below unity, with additional bin-to-bin structure, but they do not show a monotonic suppression with increasing mass or a common radius-dependent ordering across the populated range. The statistical uncertainties are bin dependent and generally become larger toward the high-mass tail; isolated ratio excursions should therefore not be over-interpreted as evidence for a systematic mass-dependent modification.

\begin{figure*}[!htbp]
  \centering
  \includegraphics[width=0.4900\linewidth]{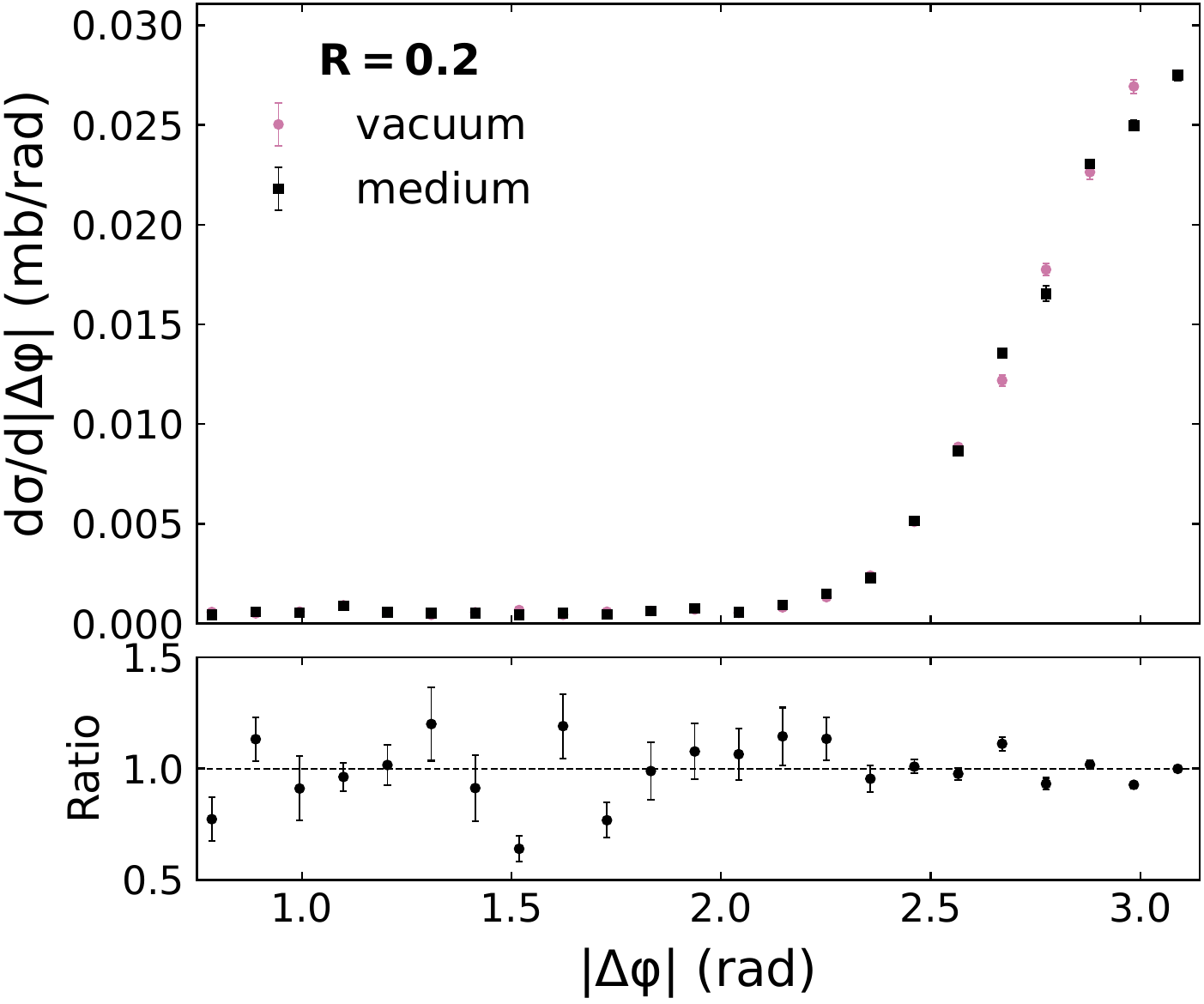}
  \includegraphics[width=0.4900\linewidth]{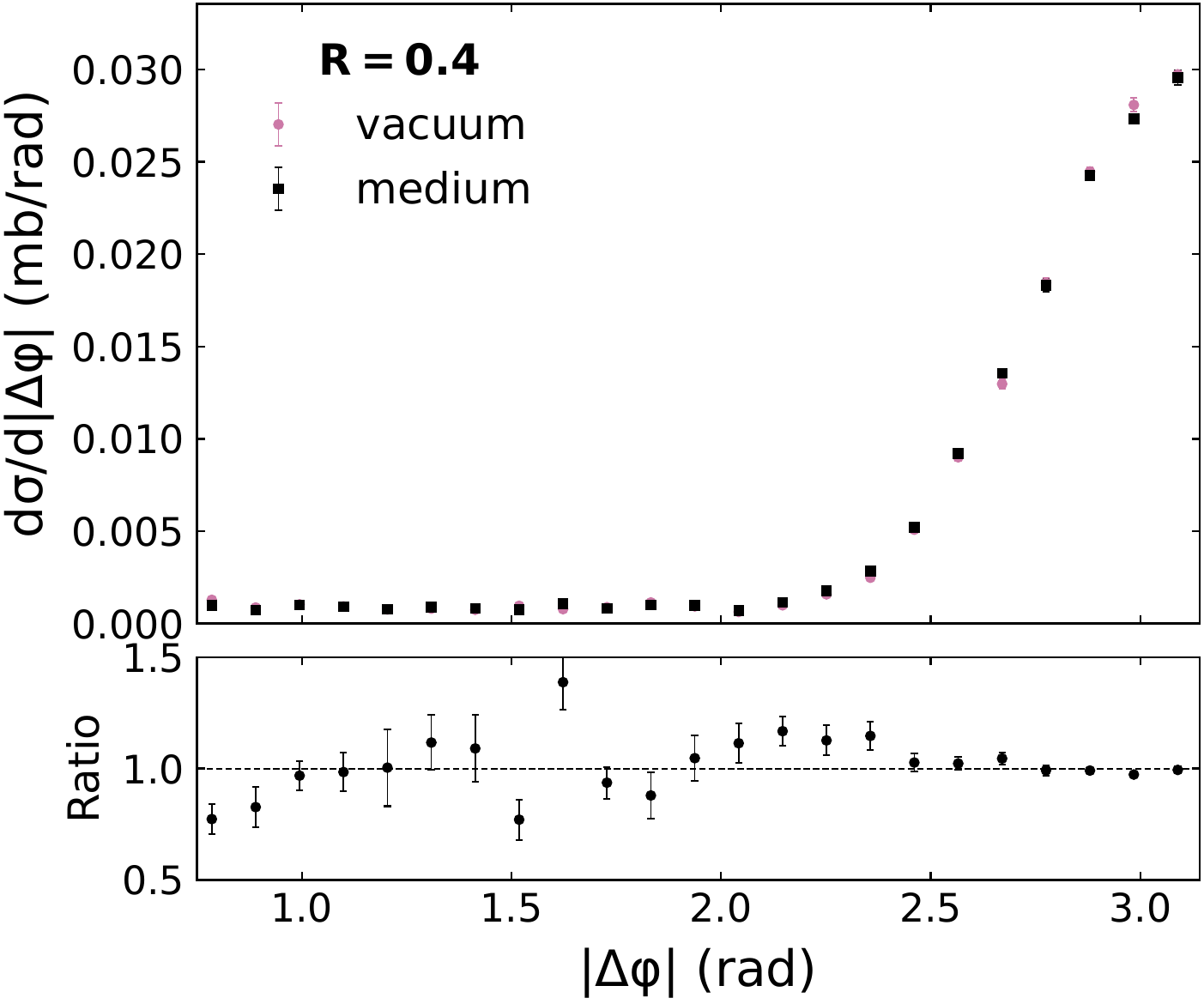}
  \caption{Folded azimuthal-separation distributions for the realistic vacuum and medium calculations in the top-20\% EA class. Jets are reconstructed with $R=0.2$ in the left panel and $R=0.4$ in the right panel. The lower panels show medium divided by vacuum.}
  \label{fig:vacmedium-dphi}
\end{figure*}

The folded azimuthal-separation distributions in Fig.~\ref{fig:vacmedium-dphi} retain the strong back-to-back peak near $\absdphi=\pi$ for both radii. Near the peak the vacuum and medium distributions remain close. For both radii, the ratio lies above unity in several bins between $\absdphi\approx2$ and 2.4, a region containing a small fraction of the dijet yield, while the ratios at smaller $\absdphi$ fluctuate strongly. Because the peak itself is not visibly depleted and the fluctuations elsewhere are of similar size, these distributions do not resolve a clear medium-induced broadening.

\begin{figure*}[!htbp]
  \centering
  \includegraphics[width=0.4862\linewidth]{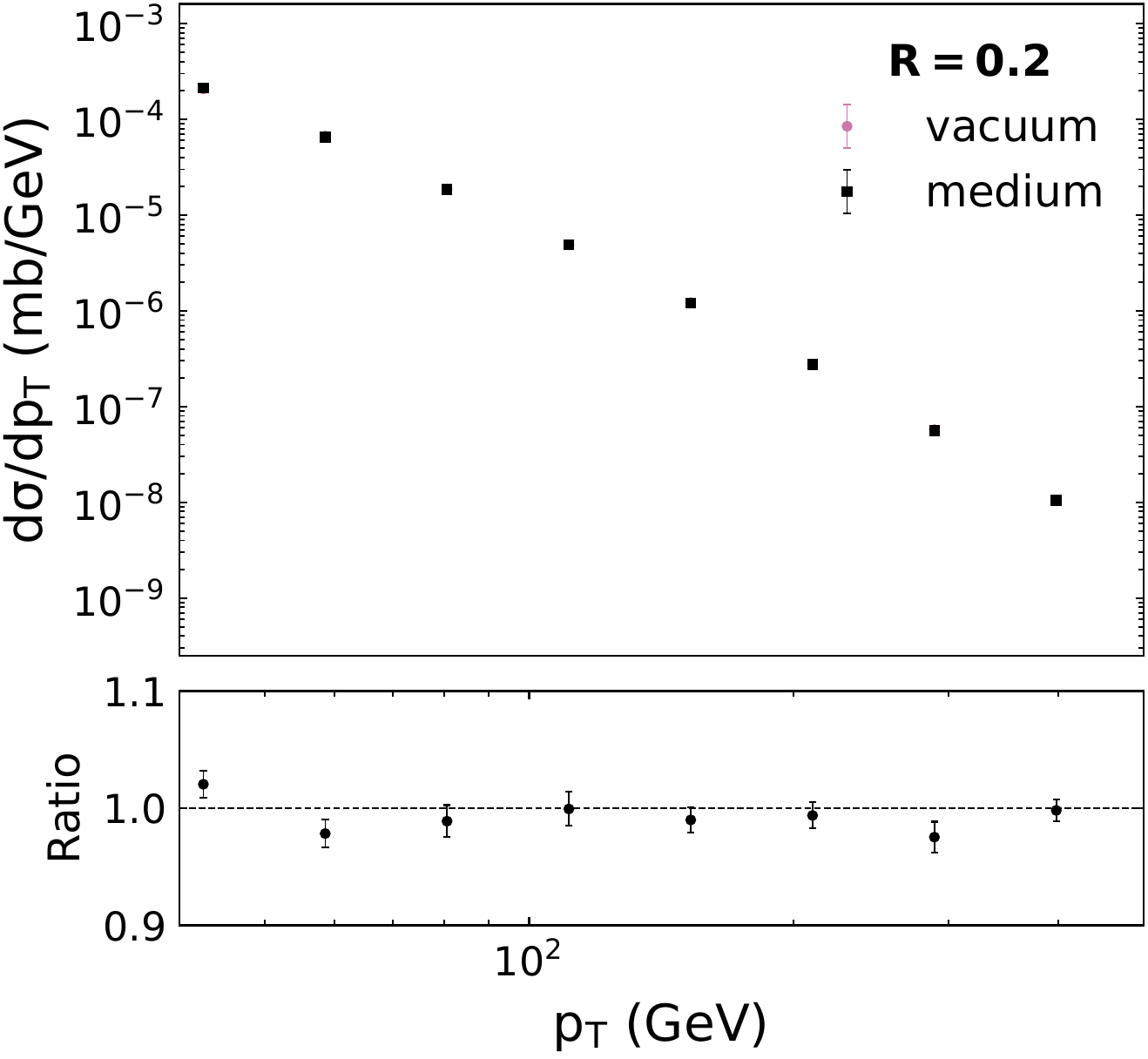}
  \includegraphics[width=0.4938\linewidth]{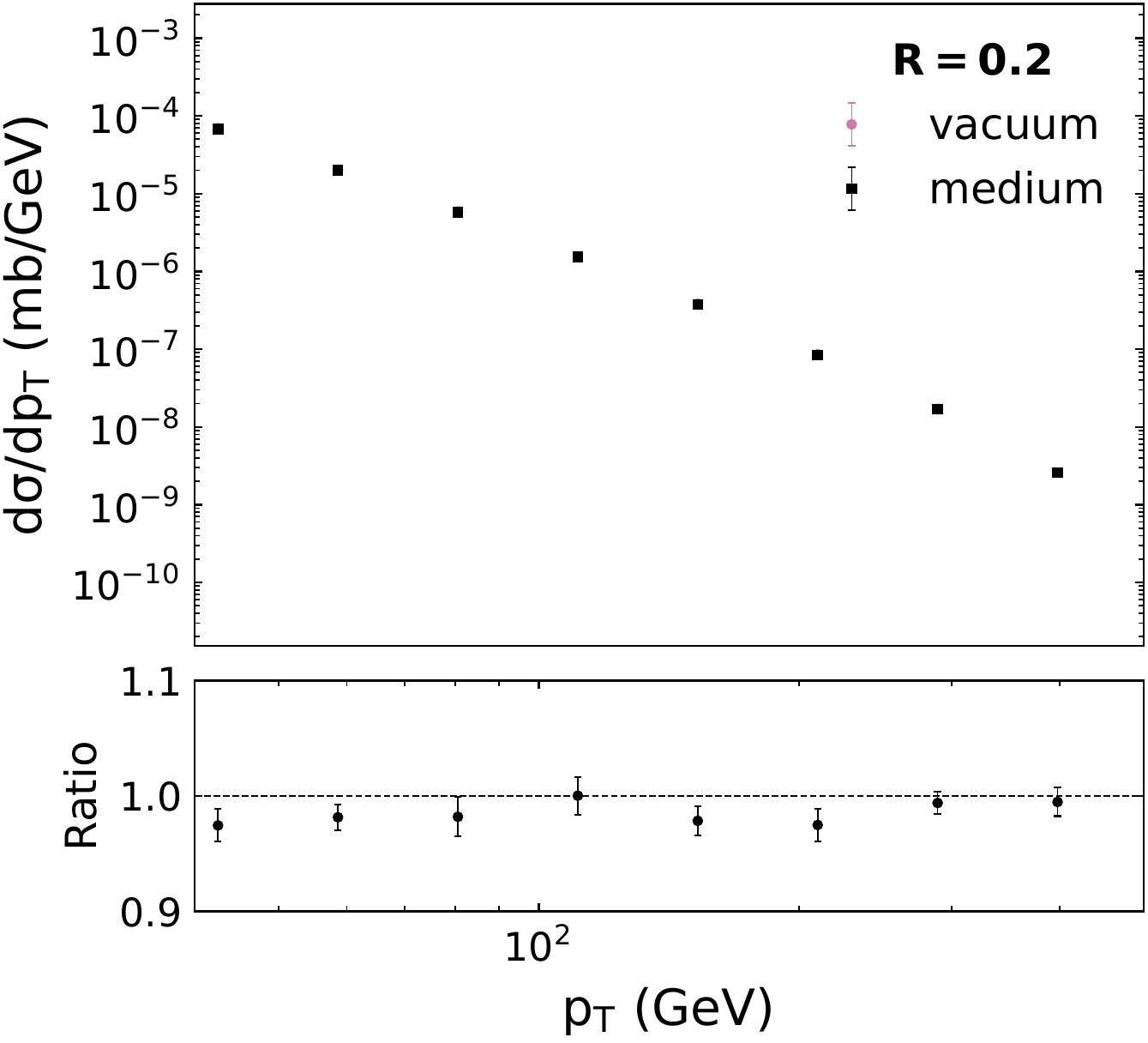}
  \caption{Dijet-selected leading-jet (left) and subleading-jet (right) $\pt$ spectra for the realistic vacuum and medium calculations in the top-20\% EA class, using $R=0.2$. The underlying dijet selection uses $10\leq\pt^{\rm lead}\leq500$~GeV. The lower panels show medium divided by vacuum.}
  \label{fig:vacmedium-jetpt}
\end{figure*}

Figure~\ref{fig:vacmedium-jetpt} compares the dijet-selected leading- and subleading-jet spectra for $R=0.2$. For both jets, the medium-to-vacuum ratio lies slightly below unity in most displayed bins, with a mild non-monotonic dependence on $p_{\rm T}$. The size of this reduction does not change systematically with $p_{\rm T}$ across the plotted range.

\begin{figure*}[!htbp]
  \centering
  \includegraphics[width=0.4803\linewidth]{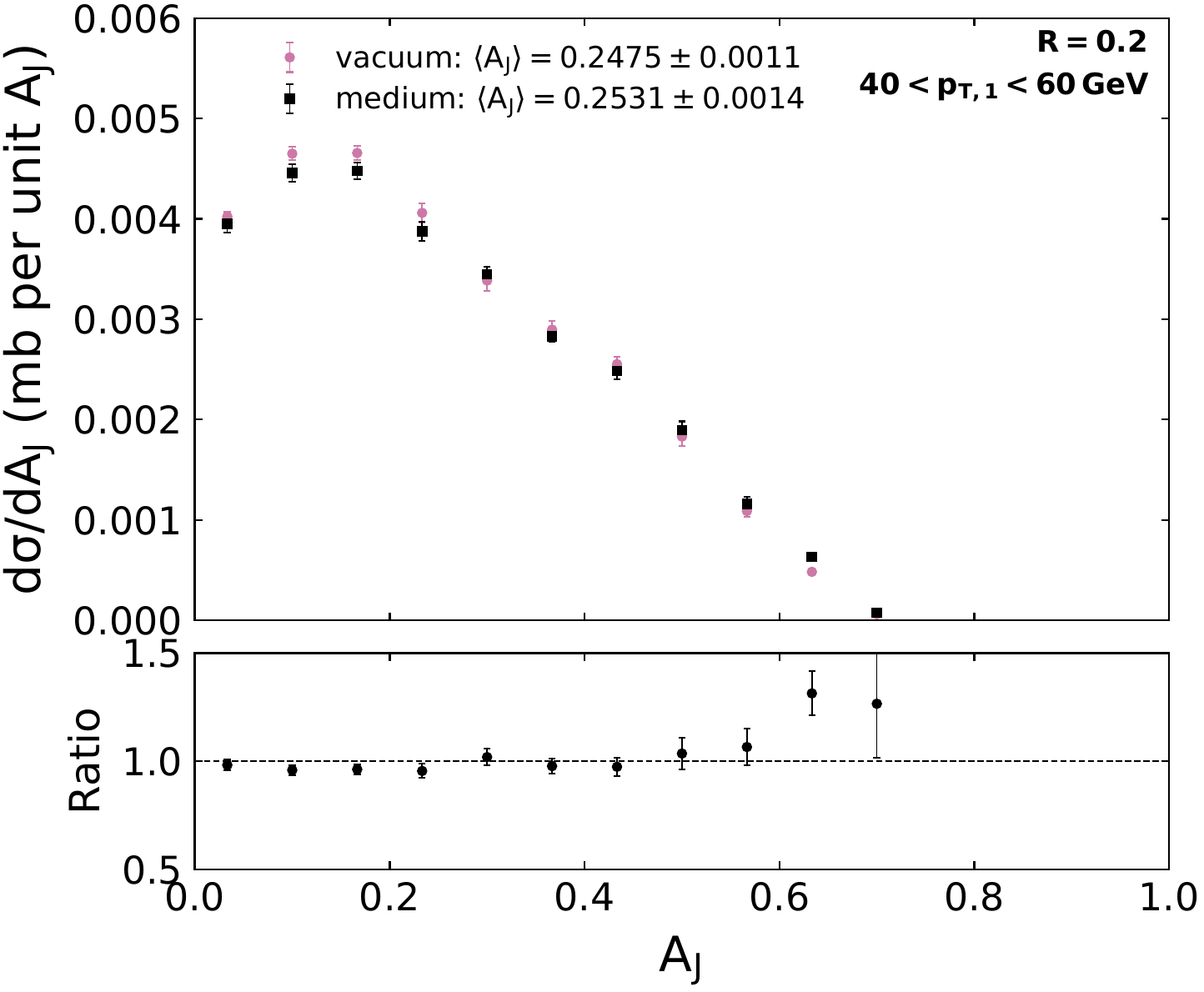}
  \includegraphics[width=0.4997\linewidth]{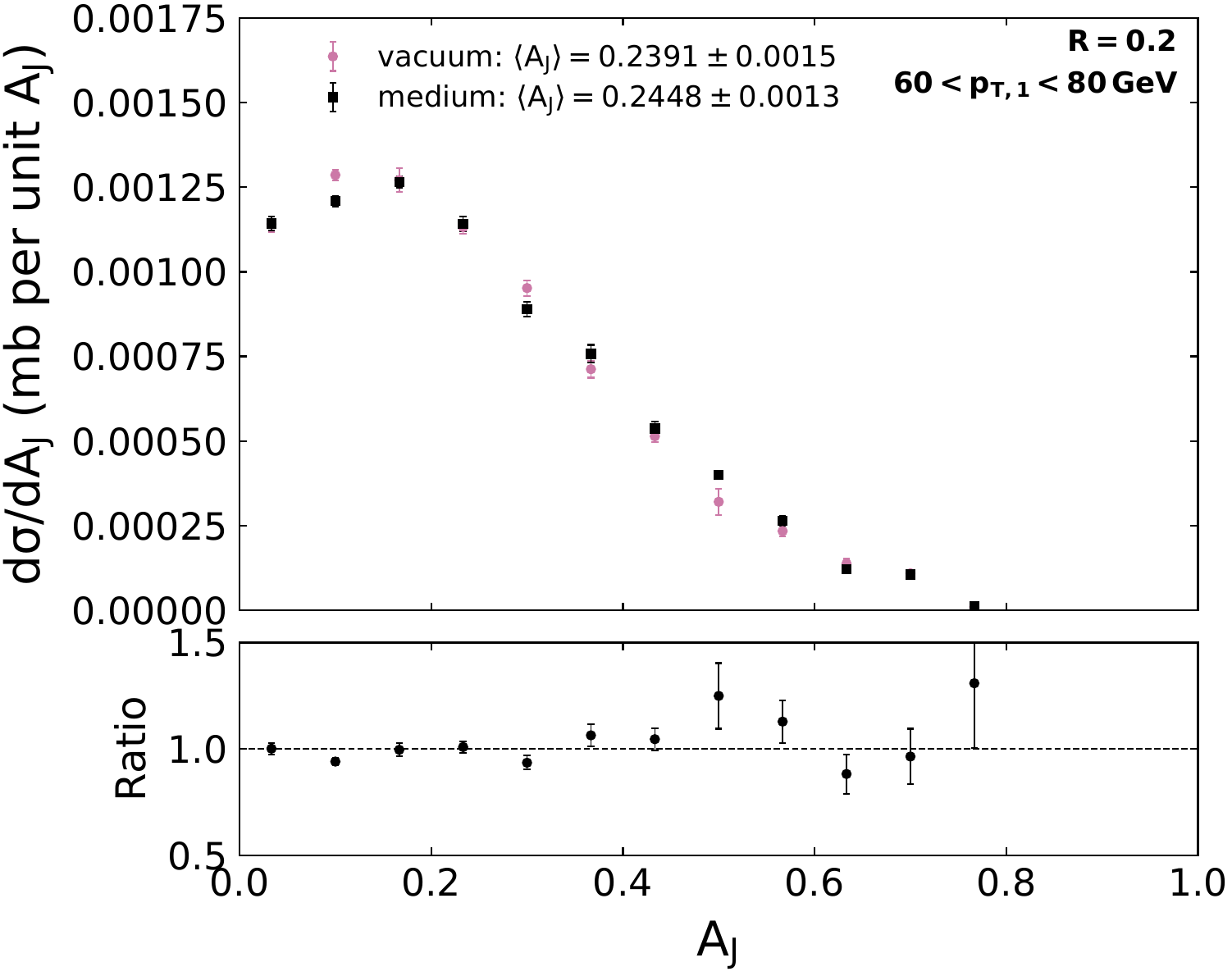}
  \caption{Dijet-asymmetry distributions for the realistic vacuum and medium calculations in the top-20\% EA class with $R=0.2$. The left panel shows $40<\pt^{\rm lead}<60$~GeV, and the right panel shows $60<\pt^{\rm lead}<80$~GeV. The lower panels show medium divided by vacuum.}
  \label{fig:vacmedium-aj}
\end{figure*}

The dijet-asymmetry distributions are shown in Fig.~\ref{fig:vacmedium-aj}. For $40<\pt^{\rm lead}<60$~GeV, the mean increases from $0.2475\pm0.0011$ in vacuum to $0.2531\pm0.0014$ in the medium calculation, a shift of $0.0056\pm0.0018$ (${\approx}3.1\sigma$). For $60<\pt^{\rm lead}<80$~GeV, the mean increases from $0.2391\pm0.0015$ to $0.2448\pm0.0013$, a shift of $0.0057\pm0.0020$ (${\approx}2.9\sigma$). The distributions nevertheless overlap strongly over most of the measured $\aj$ range, and in both intervals the medium excess appears mainly at large $\aj$.

\begin{figure*}[!htbp]
  \centering
  \includegraphics[width=0.4850\linewidth]{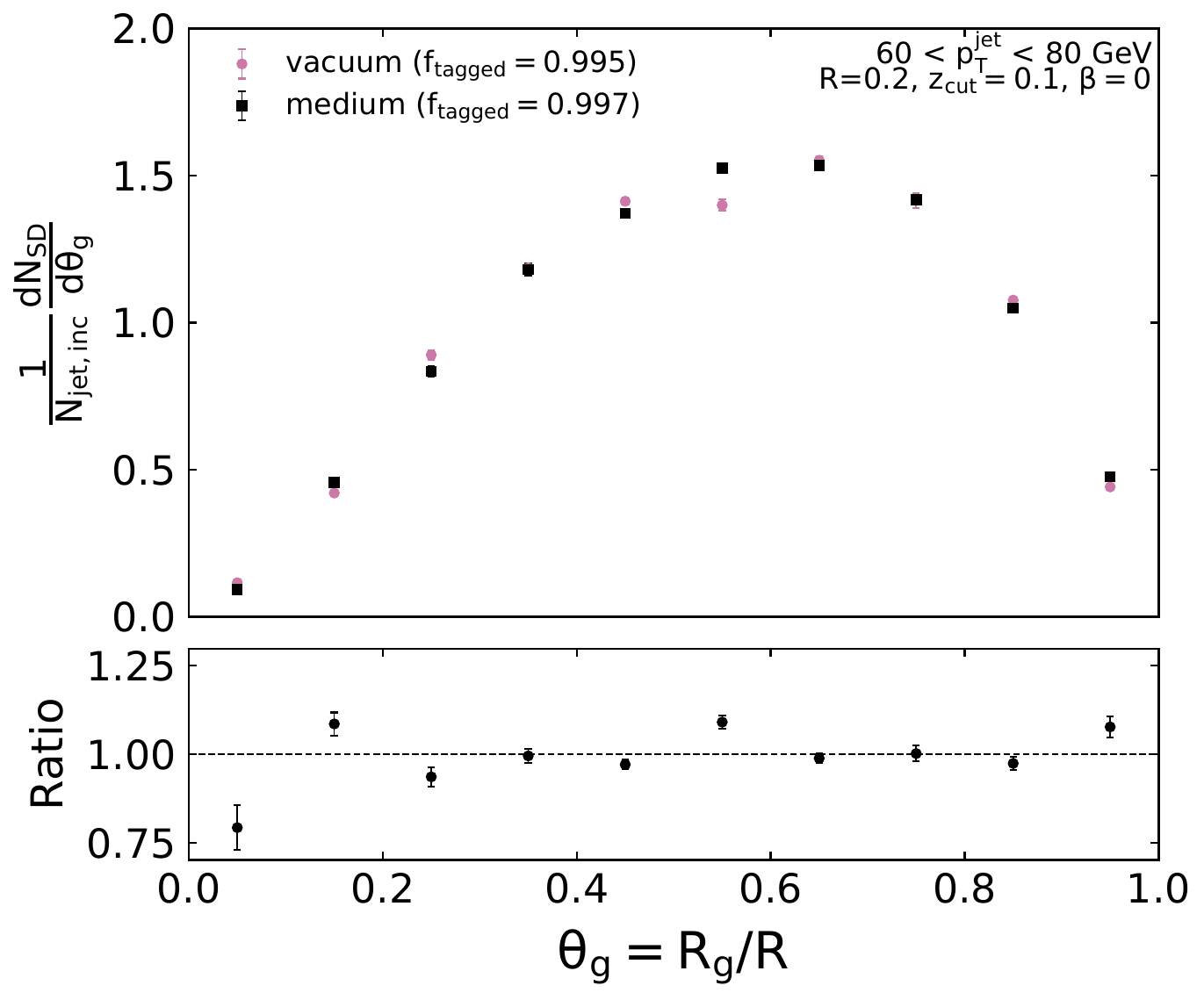}
  \includegraphics[width=0.4950\linewidth]{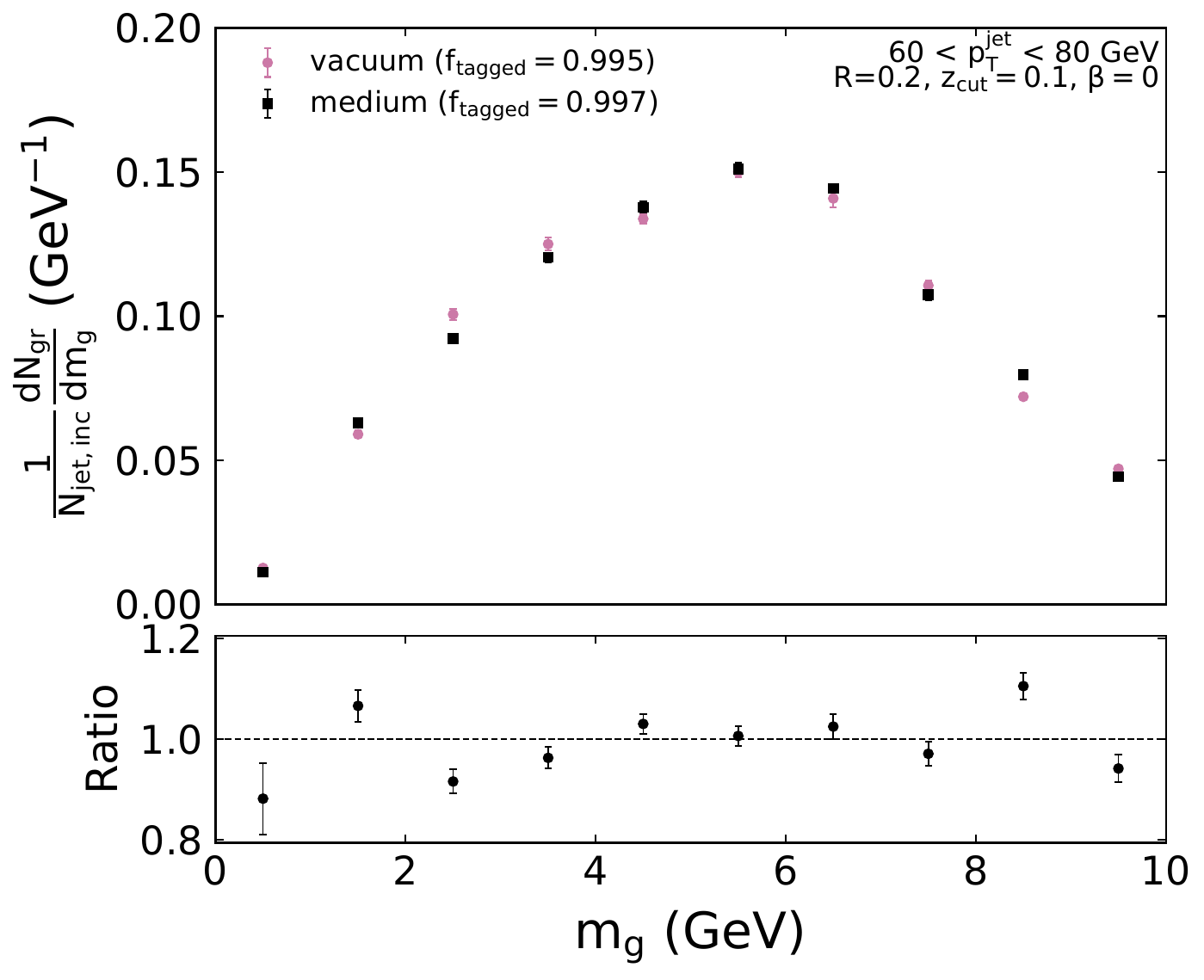}
  \caption{Soft-Drop groomed observables for inclusive jets with $60<\pt^{\rm jet}<80$~GeV in the top-20\% EA class, using $R=0.2$. The left panel shows the scaled groomed radius $\theta_{\rm g}=R_{\rm g}/R$, and the right panel shows the groomed mass $m_{\rm g}$. Soft Drop is applied with $z_{\rm cut}=0.1$ and $\beta=0$. The spectra are normalized to the weighted inclusive parent-jet yield in the same $R$ and $\pt$ interval. The lower panels show medium divided by vacuum.}
  \label{fig:vacmedium-groomed}
\end{figure*}

Figure~\ref{fig:vacmedium-groomed} shows the Soft-Drop groomed-radius and groomed-mass distributions. The weighted Soft-Drop tagging fractions are close to unity and nearly unchanged between the two calculations: $f_{\rm tagged}=0.995$ in vacuum and $0.997$ in medium.

The $\theta_{\rm g}$ distribution in Fig.~\ref{fig:vacmedium-groomed} contains localized vacuum--medium differences, but the ratio alternates above and below unity across the populated range. It does not show the sustained enhancement at small $\theta_{\rm g}$ together with suppression at large $\theta_{\rm g}$ that would constitute a clear angular-narrowing pattern.

The groomed-mass distribution in Fig.~\ref{fig:vacmedium-groomed} likewise remains close between the two samples over the densely populated part of the distribution. The ratio contains localized bin-to-bin departures, while the highest-mass portion becomes progressively less populated. Across the displayed range there is no monotonic transfer of yield from high to low $m_{\rm g}$, or the reverse, and therefore no single groomed-mass redistribution pattern is resolved.

\section{Conclusion}
\label{sec:conclusion}

We have examined the response of jet and dijet observables to medium-induced effects in pp collisions at $\sqrts=13$ TeV using a controlled Bjorken-expanding brick and a realistic event-by-event small-system calculation.

The brick calculation identifies the dijet invariant mass as the most responsive observable: in the well-populated region its ratio to the vacuum reference decreases with brick length, more strongly for $\qhatnought=0.50$~GeV$^2$/fm and for $R=0.2$. The jet spectra respond more weakly, most visibly for the subleading jet and the narrower radius; $\langle\aj\rangle$ increases only slightly, and no azimuthal broadening is resolved.

In the realistic top-20\% EA class, the ungroomed medium-to-vacuum ratios lie  slightly below unity without a monotonic kinematic dependence, and $\langle\aj\rangle$ increases by more than its statistical uncertainty in both leading-jet intervals, $40<\pt^{\rm lead}<60$ and $60<\pt^{\rm lead}<80$~GeV. Neither an angular narrowing of $\theta_{\rm g}$ nor a systematic redistribution of $m_{\rm g}$ is resolved.

The brick thus shows that these observables respond in an ordered way to the medium path length and transport strength, whereas the realistic calculation yields a small reduction of the dijet yield with observable-specific structure rather than a common ordered pattern. Within the present model setup and statistical precision, no single observable in the realistic sample exhibits a modification that by itself would establish jet quenching in high-activity pp collisions, and this study makes no claim in favor of or against quark--gluon plasma formation in pp collisions. The results instead set the scale of the expected medium-induced signal and provide a basis for more differential comparisons and future studies with increased statistics or alternative small-system medium scenarios.

The medium-induced modifications found in this work are small compared with the multiplicity-dependent changes of jet observables that ALICE has already measured in pp collisions at $\sqrts=13$~TeV and that PYTHIA reproduces qualitatively without a medium: the jet yield rises with multiplicity~\cite{ALICE:2022JetMultiplicity}, jet fragmentation becomes softer~\cite{ALICE:2024IntraJet}, and recoil-jet acoplanarity broadens~\cite{ALICE:2024HMppAcoplanarity}. Because no medium-free reference exists at the same event activity, a measurement of the observables examined in this work would have to separate a medium-induced signal of this size from such selection effects, using a reference that reproduces them, such as event generators or lower-multiplicity event classes.

\begin{acknowledgments}
The authors acknowledge the Wayne State University's Computing \& Information Technology (C\&IT) High Performance Computing Services for access to the Wayne State Grid and the computational resources used in this work. L.F. acknowledges support from the Thomas C. Rumble University Graduate Fellowship at Wayne State University. This work was supported in part by the United States Department of Energy, Office of Nuclear Physics, under grant Nos. DE-FG02-92ER40713 and DE-SC0013460. Anthropic's Claude (Opus 5.5) was used to assist with debugging and reviewing the simulation, analysis, and plotting code. All physics choices, code changes, and numerical results were independently verified by the authors, who retain full responsibility for the work.
\end{acknowledgments}

\onecolumngrid
\section*{Data Availability}
The configuration files, custom analysis and plotting codes, as well as the final histogram files supporting the reported results, are available from the authors upon reasonable request. The generated event samples are not publicly deposited because of their very large size.
\twocolumngrid

\clearpage
\onecolumngrid
\appendix

\section{Configuration and statistical details}
\label{app:configuration-statistics}

This appendix summarizes the detailed physics parameters of the controlled and realistic calculations and the statistical conventions used for the final distributions.

\newcommand{\parameterrow}[2]{%
\parbox[t]{0.36\linewidth}{\raggedright #1\strut} & \parbox[t]{0.58\linewidth}{\raggedright #2\strut} \\[0.18em]}
\newcommand{\tablesection}[1]{\multicolumn{2}{@{}l}{\rule{0pt}{1.2em}\textit{#1}} \\[0.18em]}

\subsection{Controlled vacuum--brick calculation}
\label{app:controlled-parameters}

\begin{table}[!htbp]
\caption{Parameters of the controlled vacuum--brick calculation, following the PP19 prescription of Ref.~\cite{JETSCAPE:2020ppResults}. Entries apply to both samples unless stated otherwise.}
\label{tab:controlled-parameters}
\centering
\small
\begin{tabular}{@{}ll@{}}
\toprule
\parameterrow{\textbf{Parameter}}{\textbf{Value}}
\midrule
\tablesection{Hard process}
\parameterrow{Collision system}{pp, $\sqrts=13$~TeV}
\parameterrow{Hard-scattering range}{$30<\pthat<3000$~GeV}
\parameterrow{PYTHIA~8 processes}{hard QCD; prompt photons off}
\parameterrow{Initial-state radiation and MPI}{PYTHIA~8, on}
\parameterrow{Final-state radiation}{MATTER (PYTHIA~8 final-state shower off)}
\parameterrow{Parton $\pt$ threshold for MATTER}{2~GeV}
\tablesection{MATTER shower}
\parameterrow{Switching scale}{$Q_0=1$~GeV}
\parameterrow{Initial virtuality}{$Q_{\rm ini}=p_{\rm T}/2$}
\parameterrow{Coupling}{$\alpha_s=0.25$}
\parameterrow{Additional broadening}{off}
\tablesection{Brick medium (brick samples only)}
\parameterrow{Evolution}{transversely uniform, Bjorken longitudinal expansion}
\parameterrow{Initial conditions}{$\tau_0=0.2$~fm/$c$, $T_0=0.35$~GeV}
\parameterrow{Cutoff temperature}{$T_c=0.155$~GeV}
\parameterrow{Brick length}{$L=0.5$, 1.0, 2.0~fm}
\parameterrow{Transport coefficient}{constant $\qhat_0=0.25$, 0.50~GeV$^2$/fm}
\parameterrow{Medium recoil}{on; recoil partons hadronized, backreaction subtracted in jet clustering (Sec.~\ref{sec:observables})}
\tablesection{Vacuum reference}
\parameterrow{Medium}{none ($\qhat=0$, recoil off)}
\tablesection{Hadronization}
\parameterrow{Model}{colorless string hadronization with beam remnants for $\sqrts=13$~TeV}
\bottomrule
\end{tabular}
\end{table}

\clearpage
\subsection{Realistic vacuum--medium calculation}
\label{app:realistic-parameters}

\begin{table}[!htbp]
\caption{Hard-process and initial-state parameters of the realistic calculation, based on the X-SCAPE configuration of Ref.~\cite{Soudi:2024XSCAPE}. The vacuum and medium samples use identical settings.}
\label{tab:realistic-parameters-a}
\centering
\small
\begin{tabular}{@{}ll@{}}
\toprule
\parameterrow{\textbf{Parameter}}{\textbf{Value}}
\midrule
\tablesection{Hard process}
\parameterrow{Collision system}{pp, $\sqrts=13$~TeV}
\parameterrow{Hard-scattering range}{$30<\pthat<3000$~GeV}
\parameterrow{PYTHIA~8 processes}{hard QCD and prompt photons}
\parameterrow{Multiparton interactions}{on; minimum $\pt=8$~GeV}
\parameterrow{Reference $\alpha_s$}{0.15 (hard process and MPI)}
\parameterrow{Phase-space bias}{$(\pthat/4~{\rm GeV})^4$, compensated by event weights}
\parameterrow{Initial-state radiation}{iMATTER backward evolution (PYTHIA~8 initial-state shower off)}
\tablesection{Initial state}
\parameterrow{Initial condition}{three-dimensional Monte Carlo Glauber model~\cite{Shen:2020Geometry3D}}
\parameterrow{Mean rapidity loss}{1.60, 2.15, 2.15, 2.15 at initial rapidity 2, 4, 6, 10; fluctuation parameter 0.5}
\parameterrow{Source widths}{$\sigma_x=0.5$~fm, $\sigma_\eta=0.5$}
\parameterrow{String-source pre-equilibrium flow factor}{0.15}
\parameterrow{Hydrodynamic starting time}{$\tau_0=0.6$~fm/$c$}
\bottomrule
\end{tabular}
\end{table}

\begin{table}[!htbp]
\caption{Hydrodynamic, particlization, shower, and hadronization parameters of the realistic calculation. Entries apply to both samples unless marked as medium only.}
\label{tab:realistic-parameters-b}
\centering
\small
\begin{tabular}{@{}ll@{}}
\toprule
\parameterrow{\textbf{Parameter}}{\textbf{Value}}
\midrule
\tablesection{Hydrodynamics (MUSIC)}
\parameterrow{Equation of state}{NEoS-BQS~\cite{Monnai:2019NEoS}}
\parameterrow{Shear viscosity}{$\eta T/(e+P)=0.20$, constant}
\parameterrow{Bulk viscosity}{$\zeta/s$ peak 0.08 at $T=0.17$~GeV; width 0.02~GeV; asymmetry 0.1}
\parameterrow{Second-order viscous terms}{on}
\tablesection{Particlization (iSS)}
\parameterrow{Switching energy density}{$\epsilon_{\rm sw}=0.3$~GeV/fm$^3$}
\parameterrow{Samples per hypersurface}{1000; resonance decays on}
\parameterrow{Hydrodynamic events per hard event}{one (no reuse)}
\tablesection{Hard shower (MATTER and iMATTER)}
\parameterrow{Switching scale}{$Q_0=2$~GeV}
\parameterrow{Initial virtuality}{$Q_{\rm ini}=p_{\rm T}/2$}
\parameterrow{Medium coupling (medium only)}{HTL $\qhat$ with one running and one fixed coupling; $\alpha_s=0.25$}
\parameterrow{Shower--medium threshold (medium only)}{$T_c=0.16$~GeV}
\parameterrow{Recoil and additional broadening}{off}
\tablesection{Hadronization}
\parameterrow{Model}{color-connected string hadronization; final-state partons connected to the initial-state partons and Glauber beam remnants}
\bottomrule
\end{tabular}
\end{table}

\clearpage
\twocolumngrid
\subsection{Statistical uncertainties}
\label{app:statistical-conventions}

For the controlled vacuum--brick samples, all events carry unit weight. Bin uncertainties are the weighted-Poisson ($\sum w^2$) errors, combined for the absolute spectra with the PYTHIA $\sigma_{\rm gen}$ normalization uncertainty. The vacuum and brick samples are generated with independent random seeds, so the brick-to-vacuum ratio uncertainties are obtained by adding the relative uncertainties of numerator and denominator in quadrature. The quoted $\langle\aj\rangle$ uncertainties are propagated from the bin uncertainties of the corresponding $\aj$ distributions.

Statistical uncertainties for the realistic vacuum--medium comparison are evaluated with a delete-group jackknife using $N=20$ subsample groups separately within the vacuum and medium samples. For each omitted group, the observable is recomputed and the weighted highest-activity 20\% threshold is redetermined within that replica. The resulting replicas define the within-sample statistical covariance matrices. These vacuum and medium covariance matrices are then propagated independently into the medium-to-vacuum ratio; no vacuum--medium cross-sample covariance is assumed.

For the groomed observables, each replica reconstructs both the Soft-Drop-tagged numerator and the inclusive parent-jet denominator from the retained events, so the numerator--denominator correlation and the tagging-fraction uncertainty are propagated together within each sample. For the ungroomed observables, the sample cross-section-normalization uncertainty is propagated separately from the jackknife statistical uncertainty. The quoted $\aj$ means are first moments of the original analysis histograms; their statistical uncertainties are evaluated from the corresponding jackknife replicas, while the overall cross-section normalization cancels in the normalized first moments.

\renewcommand{\bibsection}{\section*{References}}
\bibliographystyle{apsrev4-1}
\bibliography{refs}

\end{document}